\documentclass[11pt,a4paper]{article}
\usepackage{amsmath,amsthm,amssymb,epsfig,latexsym}
\usepackage[numbers,square,sort&compress]{natbib}

\newcommand{\be}[1]{\begin{equation}\label{#1}}
\newcommand{\ba}[1]{\begin{multline}\label{#1}}
\newcommand{\ee}{\end{equation}}
\newcommand{\ea}{\end{eqnarray}}

\newcommand{\eq}[1]{(\ref{#1})}

\newcommand{\bra}[1]{\langle\,#1\,|}
\newcommand{\ket}[1]{|\,#1\,\rangle}

\def\a{\alpha}

\def\le{\leqslant}
\def\ge{\geqslant}

\def\sul{\sum\limits}

\def\build#1_#2^#3{\mathrel{
\mathop{\kern 0pt#1}\limits_{#2}^{#3}}}

\def\cZ{\mathcal{Z}}

\usepackage{euscript}

\def\De{\Delta}

\def\la{\lambda}

\def\sul{\sum\limits}

\def\lt({\left(}
\def\rt){\right)}

\def\det{\operatorname{det}}

\def\De{\Delta}

\def\la{\lambda}

\newtheorem{Def}{Definition}[section]

 \makeatletter
 \@addtoreset{equation}{section}
 \makeatother
 
\begin{document}

\begin{flushright}
LPENSL-TH-10/11

\end{flushright}
\par \vskip .1in \noindent

\vspace{24pt}

\begin{center}
\begin{LARGE}
\vspace*{1cm}
  {Form factor approach to the asymptotic behavior of correlation functions
in critical models }
\end{LARGE}

\vspace{50pt}

\begin{large}

{\bf N.~Kitanine}\footnote[1]{IMB, UMR 5584 du CNRS, Universit\'e
de Bourgogne, France, Nikolai.Kitanine@u-bourgogne.fr},~~
{\bf K.~K.~Kozlowski}\footnote[2]{IUPUI, Department of Mathematical Sciences, Indianapolis, USA,
 kakozlow@iupui.edu},~~
{\bf J.~M.~Maillet}\footnote[3]{ Laboratoire de Physique, UMR 5672
du CNRS, ENS Lyon,  France,
 maillet@ens-lyon.fr},\\
{\bf N.~A.~Slavnov}\footnote[4]{ Steklov Mathematical Institute,
Moscow, Russia, nslavnov@mi.ras.ru},~~
{\bf V.~Terras}\footnote[5]{ Laboratoire de Physique, UMR 5672 du
CNRS, ENS Lyon,  France, veronique.terras@ens-lyon.fr}
\par

\end{large}

\vspace{80pt}

\centerline{\bf Abstract} \vspace{1cm}
\parbox{12cm}{\small   We propose a form factor approach for the computation of the large distance asymptotic
behavior of correlation functions in quantum critical (integrable) models.
In the large distance regime we reduce the summation over
all excited states to one over the particle/hole excitations lying on the Fermi surface in the thermodynamic limit. We compute
these sums, over the so-called critical form factors, exactly. Thus we obtain the leading large distance behavior of each oscillating harmonic  of the correlation function asymptotic expansion, including the corresponding amplitudes. Our method is applicable to a wide variety of integrable models and  yields precisely the results stemming
from the Luttinger-liquid approach, the conformal field theory predictions and our previous analysis of
the correlation functions from their multiple-integral representations.
We argue that our scheme applies to a general class of non-integrable quantum critical models as well.}
\end{center}

\newpage

\section{Introduction\label{INTR}}

Form factors and correlation functions are  central objects for the study of dynamical properties of models in quantum field theory and statistical mechanics. For general interacting systems, their calculation remains a fantastic challenge. In the integrable model situation in low dimension, see e.g. \cite{Bet31,Hul38,Orb58,Wal59,LieSM61,LieM66,YanY66,FadST79,Tha81,Bax82L,GauL83,BogIK93L,JimM95L,FadLH96} and references therein, considerable progress has been made during the last forty years towards the exact calculation of these dynamical quantities  \cite{Mcc68,McCTW77,SatMJ78,McCPW81,KarW78,Smi92b,CarM90,Mus92,FriMS90,KouM93,AhnDM93,JimMMN92,JimM95L,JimM96,Zam91,LukZ97,Luk99,LukZ01,LukZ01a,KitMT99,KitMT00,MaiT00,KitMST02a,KitMST05a,KitKMST09b,GohKS04,GohKS05,BooG09,BooJMST07,BooJMS09,JimMS11,JimMS11a,CauM05}. Of particular interest are the so-called critical models where the correlation functions of local operators are believed, from Luttinger liquid approach and conformal field theory (CFT) predictions, to decay as power laws in the distance  \cite{LutP75,Hal80,Hal81a,Hal81b,BelPZ84,Car84,Aff85,BloCN86,Car86,AlcBB88,WoyE87}. There are several ways to approach the computation of correlation functions in such models, either from their massive field theories counterparts (in their short distance behavior)  or from critical lattice models (looking at their long distance properties). In such limits one should recover the picture predicted by CFT, with the possible additional insight into the explicit correspondence with the physical local operators of the microscopic initial theories. The  latter information is crucial for obtaining full control of their dynamics. Indeed, while CFT is  able to predict for example the algebraic coefficients in the operator product expansion of local operators together with the values of the various critical exponents driving the power law behavior of the correlation functions, it fails to determine the (non universal) corresponding amplitudes (vacuum expectation values and form factors) which are in fact deeply rooted in the detailed microscopic interactions of the physical models at hand.

The purpose of the present article is to show that under quite general assumptions,  which are valid in the integrable model situation, it is possible to determine exactly the asymptotic  behavior of correlation functions in critical models  starting from their expansion in terms of form factors. The use of form factors in this context might appear to be a very strange strategy (see however \cite{ColIKT93,KorS90}). Indeed, as has been demonstrated \cite{Sla90,KitKMST09c,KitKMST11a}, form factors of local operators in critical models  scale to zero with a non trivial power law in the system size (or equivalently in the mass scale). This is not surprising as it just reflects the corresponding critical dimensions of the local operators (together with those of the creation operators of the excited states) considered. Hence the form factor expansion of correlation functions of local operators in critical models is  a large (infinite)  sum of small (vanishing) terms in the infinite volume limit. There it should be  the collective (volume) effect of the sum that compensates the vanishing behavior of each term of this sum. Hence, extracting this collective effect directly from the form factor series seems a priori to be a quite involved task. However, a very interesting effect revealed itself  in our recent derivation of the leading asymptotic behavior of correlation functions of the $XXZ$ chain obtained  from their re-summed multiple-integral representations \cite{KitKMST09b}. Unexpectedly, it appeared there that the exactly computed amplitudes of the leading power law decay of the spin--spin correlation function were in fact given by  properly renormalized (by a positive power of the system size) form factors of the local spin operator. The very appealing form of this answer was a strong motivation for us to develop a method for studying such correlation functions from their form factor expansion.

The method that we propose in the present paper opens up a way to analyze the large distance asymptotic behavior
of correlation functions of critical models from their form factor expansion. Namely, it
enables us to determine the leading behavior of each oscillating harmonic in the asymptotic expansion
of the correlation functions, including not only the exact power law decays but also the associated amplitudes. To describe our method, we first study  the correlation functions of quantum integrable models that are solvable by using the Bethe ansatz in their critical regime. The typical example considered in this paper is
the integrable Heisenberg  spin chain. However, the method that we describe in this paper is also applicable to the
one-dimensional Bose gas (or quantum non-linear Schr\"odinger model), and more generally to any quantum integrable model possessing determinant representations
for the form factors of local operators \cite{KorS99,KitMT99,DegM09,Oot04}. In fact we will further argue that, under quite natural assumptions,  it  applies  to a general class of non-integrable critical models as well.

Generally speaking, independently on the integrability of a quantum model, the zero temperature limit of a correlation function
of  local operators located at two different space points $x'$ and $x'+x$, say $ {\cal O}_1(x')$ and ${\cal O}_2(x'+x)$, reduces to the ground state expectation value of their product:
\begin{equation}\label{Cor-fun}
 \langle {\cal O}_1(x'){\cal O}_2(x'+x)\rangle\equiv\frac{\langle\psi_g \, |\, {\cal O}_1(x')\, {\cal O}_2(x+x')\, | \, \psi_g\rangle}
 {\langle\psi_g\,|\,\psi_g\rangle} \;.
%
 \end{equation}
Above, $|\,\psi_g\rangle$ denotes the ground state and the coordinate $x$ can be either a continuous parameter (for one dimensional field theories) or a
discrete one (for lattice models).

The form factor expansion of the zero temperature correlation function is obtained by inserting a complete set of eigenstates $\ket{\psi'}$ of the Hamiltonian between the local operators:
\begin{equation}\label{Cor-fun-ff0}
 \langle {\cal O}_1(x'){\cal O}_2(x'+x)\rangle=
 \sum_{|\,\psi'\rangle} {\cal F}^{(1)}_{\psi_g\,\psi'}(x')\; {\cal F}^{(2)}_{\psi'\,\psi_g}(x+x').
 \end{equation}
Thus the correlation function is represented as a sum of matrix elements (the so-called form factors)
 $ {\cal F}^{(j)}_{\psi_g\,\psi'}(x)$ of the local operators ${\cal O}_j$ taken between the ground state $\ket{\psi_g}$
and any excited state $\ket{\psi'}$:
\begin{equation}\label{Def-ff0}
 {\cal F}^{(1)}_{\psi_g\,\psi'}(x')=\frac{\langle\psi_g \, |\, {\cal O}_1(x') | \, \psi'\rangle}
 {\sqrt{\langle\psi_g\,|\,\psi_g\rangle\langle\psi'\,|\,\psi'\rangle}},\qquad
 {\cal F}^{(2)}_{\psi'\,\psi_g}(x+x')=\frac{\langle\psi' \, | {\cal O}_2(x+x')\, | \, \psi_g\rangle}
 {\sqrt{\langle\psi_g\,|\,\psi_g\rangle\langle\psi'\,|\,\psi'\rangle}}.
 \end{equation}
What makes the representation \eq{Cor-fun-ff0} very appealing is its straightforward  physical interpretation. Moreover, it proved to provide a very efficient way to evaluate the correlation functions numerically as soon as exact representations for the form factors are known \cite{CauM05,CauHM05,PerSCHMWA06,PerSCHMWA07,CauCS07}.

The form factor expansions are well understood in the case of massive models \cite{KarW78,Smi92b,CarM90,Mus92,FriMS90,KouM93,AhnDM93}. Actually, in these models they are extremely effective in
computing the long distance  exponential decay of the associated correlation function. In particular,  the leading long distance behavior of the correlation functions generally stems there from a few classes of excited states only. However, as we already discussed, the structure and
properties of form factor expansions  in massless (critical) models are more involved, as  an infinite tower of classes of excited states contributes to the leading asymptotic behavior of the correlation function, in contrast to the massive regime case.

There has recently  been important progress in computing the asymptotic behavior of  correlation functions for quantum integrable models; these approaches rely basically on the Riemann--Hilbert analysis of related Fredholm determinants, and are quite technical  \cite{KitKMST09a,KitKMST09b,KozT11,Koz11}. We believe that the method that we propose in this paper  is simpler as it follows a path closer to the physical intuition and hence should be of a wider use (in particular beyond the integrable case).

As we just mentioned,  for critical models, each  term of the form factor series scales to zero as some non-trivial power $\theta$ (depending on the excited state $\ket{\psi'}$ selected by the local operators considered) of
the system size $L$ \cite{KitKMST11a,KitKMST09c,Sla90} for $L \to \infty$. In translation invariant systems we have
 \begin{equation}\label{ff-struct}
 {\cal F}^{(1)}_{\psi_g\,\psi'}(x') \cdot {\cal F}^{(2)}_{\psi'\,\psi_g}(x+x')
 =L^{-\theta}\, e^{ix{\cal P}_{ex}}\,A.
 \end{equation}
Here the dependence on the distance of the form factor is contained in the phase factor $e^{i x {\cal P}_{ex}}$, where ${\cal P}_{ex}$
is the momentum of the excited state $\ket{\psi'}$ relative to the ground state. The amplitude $A$ does not
depend on $x$ and remains finite in the thermodynamic limit.

Our strategy for the analysis of the sum \eq{Cor-fun-ff0} is as follows. Initially, the sum \eq{Cor-fun-ff0} runs through all the eigenstates
and for a finite value of the distance $x$ all eigenstates should indeed be considered.
However, in the long distance regime, as follows from \eqref{ff-struct}, each term in the form factor series  is quickly oscillating with the distance. This means that, similarly to what happens for oscillatory integrals, the leading behavior of the form factor series stems from the localization of the sum around specific critical points (like saddle points or edges of the summation interval).
This fact allows us to carry out several reasonable simplifications
of the summation range that do not affect the leading asymptotic behavior of the
correlation function. First of all,  we restrict the form factor sum \eqref{Cor-fun-ff0} to the summation over excitations of one specific type only, namely,
the particle--hole excitations\footnote{%
 Hence,  in particular, bound states will not contribute in this approach, although we believe that they could appear in the analysis of dynamical correlation functions. Of course, for some models such as the non-linear Schr\"{o}dinger model,  all the excited states with finite excitation energy can be described in terms of particles and holes.}.
Second, in the large distance regime  we take into account only contributions coming from excitations close to the endpoints of the Fermi
zone, the so-called critical excited states \cite{KitKMST09b,KitKMST11a}. In this way, we naturally restrict ourselves to the excitations appearing
in the conformal part of the spectrum of the model. We will show that the resulting restricted sum over critical excited states can be computed {\em exactly}.
The asymptotic expansion for the two-point functions of the $XXZ$ model that we obtain in this way confirms the predictions
stemming from a correspondence with  CFT/Luttinger liquid approaches and agrees with the  part of the asymptotic behavior  obtained
through the exact Riemann--Hilbert methods \cite{KitKMST09a,KitKMST09b,KozMS11a,KozT11,Koz11}.

The paper is organized as follows. In  Section 2 we define the critical form factors and formulate our main strategy for evaluating their sums in the large distance limit.  We give some arguments confirming this approach. In  section 3 we perform exactly the summation  over the critical form factors. This computation is given in a general, model independent framework. It gives the leading asymptotic behavior of all the oscillating harmonics. In section 4  we apply this general result to compute the two-point functions  for the Heisenberg $XXZ$ chain. We reproduce the leading asymptotic behavior of the two-point functions as predicted by  CFT and the correlation amplitudes obtained by the
Riemann--Hilbert approach. Finally, in the conclusion, we argue that the above scheme applies to a large class of critical (integrable and non-integrable) models.  In Appendix \ref{Appendix proof summation formula}
we prove the summation formula which permits us to compute  all the above critical sums exactly.

\section{Form factor series in the large distance limit}

Let us consider the case of quantum integrable models for which the ground state solution of the Bethe
equations can be described in terms of real rapidities $\lambda_j$ densely filling (with a density $\rho(\la)$) the Fermi zone $[-q,q]$. In  such models, the logarithmic Bethe ansatz equations for the ground state Bethe roots $\la_j$
take the form \cite{Hul38,LieSM61,LieM66,Orb58,YanY66}:
 \begin{equation}\label{TBE}
 Lp_0(\la_{j})-\sum_{k=1}^{N}\vartheta(\la_{j}-\la_{k})=2\pi \left(j-
 \frac{N+1}2\right),
 \qquad j=1,\dots,{N}.
 \end{equation}
Here $L$ is the size of the model, and $N$ is the number of quasi-particles in the ground state. The functions $p_0(\lambda)$ and $\vartheta(\lambda)$ are the bare momentum and bare scattering phase of the
quasi-particles in the corresponding model. The value $N$ depends on the parameters of the model (such
as the chemical potential or external magnetic field). When we take the thermodynamic limit $L\to\infty$, the ratio $N/L$ has a finite limit which we call the total density $D=\lim_{L\to\infty}\frac NL$.

In order to describe the form factor sum one needs to have a characterization of all the excited states as well.
In the Bethe ansatz framework, the excited states are parameterized by rapidities $\{\mu_{\ell_a}\}_1^{N'}$
solving a set of equations similar to \eqref{TBE} but involving other choices of integers
$\ell_1,\dots,\ell_{N^{\prime}}$ in the $rhs$:
 \begin{equation}\label{TBE-lj}
 Lp_0(\mu_{\ell_j})-\sum_{k=1}^{N'}\vartheta(\mu_{\ell_j}-\mu_{\ell_k})=2\pi \left(\ell_j-
 \frac{N'+1}2\right),
 \qquad j=1,\dots,{N'}.
 \end{equation}

Above, $N'$ corresponds to the number of quasi-particles contained in the given excited state.
The values of $N'$  to be considered for carrying out the form factor sums depend on the type of local operator
${\cal O}$ that one deals with.
A local operator $ {\cal O}$ only connects states having a `small' difference in the number of quasi-particles.
In other words, for a given local operator, there exists an integer $k$ such that only states with $|N-N'|\le k$ give rise to non-zero
form factors. For instance, later on, we will focus on the  $XXZ$ spin-$1/2$ chain.
There, we will consider the local spin operators, where either $N'=N$ or $N'=N\pm 1$.
We do however stress that our method allows one to consider more complicated cases of local operators that connect more
distant values of $N$ and $N'$.

For  excited states of the type that we are interested in, it is convenient to describe the solutions of the Bethe equations in terms of a set of integers  alternative to $\ell_1,\dots,\ell_{N'}$. Namely one characterizes the excitations in terms of holes
in the Dirac sea of integers for the ground state $\ell_j=j$ and additional, particle-like integers outside of the zone
$\{1,\dots, N' \}$.  In other words, the integers $\ell_j$ describing an excited $n$-particle--hole state take the form
\begin{equation}\label{label integers}
\ell_a =a \  , \; a \in \{ 1,\dots, N'\} \setminus \{h_1,\dots, h_n\} \qquad
\ell_{h_a} = p_a  \ , \; p_a \in \mathbb{Z} \setminus \{1,\dots, N' \} .
\end{equation}
Each choice of pairwise distinct quantum numbers $\{p_a\}$ and $\{h_a\}$  characterizes an excited state associated with a configuration of the particle  and  the hole rapidities  $\{\mu_{p_a}\}$ and $\{\mu_{h_a}\}$, see e.g., \cite{GauL83, BogIK93L}.

The particle/hole excitation contribution to the two-point function has the form
\begin{equation}\label{Cor-fun-ff}
 \langle {\cal O}_1(x')\,{ \cal O}_2(x+x')\rangle_{ph}=\lim_{L\to\infty}
 \sum_{\{p\},\{h\}} L^{-\theta}\, e^{ i x {\cal P}_{ex} }
\,A(\{\mu_p\},\{\mu_h\}|\{p\},\{h\}).
 \end{equation}
Here we have explicitly indicated that the amplitude $A$ \eqref{ff-struct} depends
on the particle/hole rapidities $\{\mu_p\}$ and $\{\mu_h\}$ as well as on the
corresponding integer quantum numbers $\{p\}$ and $\{h\}$.  Let us comment briefly on this dual dependence (see \cite{KitKMST11a} for more details).
If all particle/hole rapidities are separated from the Fermi boundaries, then the amplitude
smoothly depends only on these quantities $A=A(\{\mu_p\},\{\mu_h\})$. However, as soon as these rapidities approach the Fermi boundaries, the amplitude shows a discrete structure:   a microscopic (of order $1/L$) deviation of a particle (hole) rapidity, corresponding to a finite shift of the integer quantum numbers $\{p\}$ or $\{h\}$,   leads to a macroscopic change of $A$. Hence, in this case, one should take into account the discrete structure of the amplitude, and therefore we write $A=A(\{\mu_p\},\{\mu_h\}|\{p\},\{h\})$.

In the following, we will argue that the oscillatory character of the sum in \eqref{Cor-fun-ff} localizes it, in the absence of any saddle points of the oscillating exponent\footnote{Such saddle points could appear in the time dependent case and will be considered in a forthcoming publication.},  in the vicinity of the Fermi boundaries $\pm q$. For this purpose, we introduce several definitions \cite{KitKMST11a}.

\begin{Def}
An $n$ particle--hole excited state $\{ \mu_{\ell_a} \}$ is called a critical
excited state if the rapidities $\{ \mu_{p_a} \}$ and $\{ \mu_{h_a} \}$ defined by such a state
accumulate on the two endpoints of the Fermi zone in the thermodynamic limit.
Form factors corresponding to any such a state are called critical form factors.
\end{Def}

The critical excited states are characterized by the specific distribution of particles
and holes on the Fermi boundaries, $\pm q$. Namely, assume that, in the thermodynamic limit, there are $n^{\pm}_p$
particles whose rapidities are equal to $\pm q$ and $n^{\pm}_h$
holes whose rapidities are equal to $\pm q$. Then, obviously,
 \begin{equation}\label{cond-sum}
 n^+_p+n^-_p=n^+_h+n^-_h=n.
 \end{equation}

\begin{Def}
A given critical excited state belongs to the $\mathbf{P}_{\ell}$ class if the distribution of particles and holes
on the Fermi boundaries is such that
 \begin{equation}\label{cond-dif}
 n^+_p-n^+_h=n^-_h-n^-_p=\ell, \qquad -n\le\ell\le n.
 \end{equation}
Then such an excited state has momentum  $2\ell k_{{}_F}$ in the thermodynamic limit, where $k_{{}_F} = \pi D$ is the Fermi momentum. Its associated critical form factor will  also
be said to belong to the  $\mathbf{P}_\ell$ class.
\end{Def}

Now we would like to argue in favor of the localization of the form factor sum on the Fermi boundaries in the
large distance limit. For this purpose we will use an analogy with oscillatory integrals. Hence,
suppose that we deal with some multiple integral of the type
 \begin{equation}\label{int-example}
 I_n(x)=\int\limits_{\mathbb{R}\setminus [-q,q]}\,d^n\mu_{p}
 \int\limits_{[-q,q]} \,d^n\mu_{h}\;f(\{\mu_p\},\{\mu_h\})\prod_{j=1}^n
 e^{ix(p(\mu_{p_j})-p(\mu_{h_j}))},\qquad
 x\to\infty,
 \end{equation}
where $f(\{\mu_p\},\{\mu_h\})$ is a holomorphic function  in a neighborhood of the real axis and $p(\mu)$ has no saddle points on the
integration contours. By deforming the integration contours into the complex plane, one makes the integrand exponentially small everywhere
except in the vicinity of the endpoints $\pm q$.
Then the large $x$ asymptotic analysis of \eqref{int-example} reduces to the calculation of the integral in small vicinities of the
endpoints, where $f(\{\mu_p\},\{\mu_h\})$ can be replaced by $f(\{\pm q\},\{\pm q\})$.
Indeed, such replacement does not alter the leading asymptotic behavior of $I_n(x)$.
Now, if $f(\{\mu_p\},\{\mu_h\})$ has integrable singularities at $\pm q$,
for example $f(\{\mu_p\},\{\mu_h\})= (q-\mu_{h_1})^{\nu_+}(\mu_{h_1}+q)^{\nu_-}f_{reg}(\{\mu_p\},\{\mu_h\})$,
then, when carrying out the asymptotic analysis in the vicinities of $\pm q$, one has to keep the singular factors
$(q\mp\mu_{h_1})^{\nu_\pm}$ as they are, but we can replace the regular part $f_{reg}(\{\mu_p\},\{\mu_h\})$ by an appropriate constant
$f_{reg}(\{\pm q\},\{\pm q\})$. Again, this approximation does not alter the leading asymptotic behavior of $I_n(x)$.

Thus, working by analogy, we may fairly expect that, as soon as ${\cal P}_{ex}$ has no saddle point, the sum  \eq{Cor-fun-ff} will
localize on the endpoints of the summation for the integers $p_a$ and $h_a$, namely, in the language of rapidities, on the two endpoints of
the Fermi zone. In other words, the sum will localize on the critical excited states\footnote{Note that for discrete sums, the integration by part procedure could be carried out, leading in simple cases to such a result in a perfectly controlled way.}. The part of the amplitude $A(\{\mu_p\},\{\mu_h\}|\{p\},\{h\})$  smoothly depending  on the rapidities  plays the role of the regular part $f_{reg}$ in the multiple integrals \eq{int-example}. There one can
set $\{\mu_p\}$ and $\{\mu_h\}$ equal to their values in the given ${\bf P}_{\ell}$ class:
\begin{equation}\label{restriqt-q}
A(\{\mu_p\},\{\mu_h\}|\{p\},\{h\})\longrightarrow
A(\{q\}_{n_p^+}\cup \{-q\}_{n_p^-},\{q\}_{n_h^+}\cup \{-q\}_{n_h^-}|\{p\},\{h\})=
A^{(\ell)}(\{p\},\{h\}).
\end{equation}
Such replacement should not change the leading asymptotic behavior over $x$ of the form factor series.

On the other hand the remaining part of the amplitude $A^{(\ell)}(\{p\},\{h\})$ explicitly depending on $\ell$ and on the quantum numbers $p$ and $h$
plays the role of the singular factors   $(q\mp\mu_{h_1})^{\nu_\pm}$ in the  integral \eq{int-example}
since, for large system size $L$, it varies quickly  in the vicinity of the Fermi boundaries. In fact, in the thermodynamic limit, this rapid variation of the amplitude as a function of the rapidities manifests itself as a kinematical pole of the form factor whenever one particle and one hole are approaching simultaneously the same Fermi boundary. In terms of the quantum numbers $\{p\}$ and $\{h\}$ however, there is no singularity and the coefficients $A^{(\ell)}$ are always well defined.
So, fixing the  $\mathbf{P}_\ell$ class of critical form factors, we should take the  sum involving  $A^{(\ell)}(\{p\},\{h\})$ over all the excited states (namely over all the possible quantum numbers $\{p\}$ and $\{h\}$) within this class {\em before taking the thermodynamic limit}.

This picture fits perfectly into the scheme of approximations used in the application of CFT (or Luttinger liquid)
methods to the computation of the asymptotic behavior of correlation functions. Moreover, it gives  a clear interpretation of why
these predictions actually work and are so universal.

Let us point here an important remark. Although we just used the multiple-integral asymptotic behavior analogy, the sum that we consider here cannot be replaced straightforwardly by an integral sum even for  large system size $L$. The non-integer nature of the critical exponent $\theta$ would make such a correspondence quite subtle, eventually producing divergent integrals times remaining vanishing factors of $L$. This feature explains the difficulties already noted in the literature, see e.g.  \cite{LesSS96,LesS97,LesKS2003}, while trying to use the form factor approach directly in the continuum limit for massless models. The main advantage of our approach (in comparison to a field theory framework) is in computing exactly the form factor sum for $L$ large but finite, hence producing an explicit compensation for the vanishing factor $L^{-\theta}$ and making the thermodynamic limit $L\to\infty$ straightforward, as we will show in section~\ref{SCFF}.

\section{Summation of  critical form factors\label{SCFF}}

We have already insisted that critical states are gathered into various classes having distinct values of their excitation momentum
in the thermodynamic limit. More precisely, all states belonging to the $\mathbf{P}_\ell$ class
have momentum $2\ell k_{{}_F}$.
Also, we stress that all critical states
have, in the thermodynamic limit, a vanishing  excitation energy (the latter goes to zero as $1/L$),
this independently of their class.

Actually, similarly to what happens for  the excitation energy, for $L$ large but finite, the momentum of any critical state deviates
from $2\ell k_{{}_F}$ by terms of  order $1/L$. Namely, it follows from the Bethe equations \eqref{TBE} and \eq{TBE-lj} that
 \begin{equation}\label{ex-Mom}
 {\cal P}_{ex}\equiv \sum_{j=1}^{N'}p_0(\mu_{\ell_j})-\sum_{j=1}^N p_0(\lambda_j)=
\frac{2\pi}{L}\sum_{k=1}^n(p_k-h_k),
\end{equation}
where $p$ and $h$ are the  integer quantum numbers of particles and holes. It is convenient
to re-parameterize these  integers for the critical excited states as follows
 \begin{equation}\label{spec-p0}
 \begin{array}{ll}
 p_j=p_j^++N',&\mbox{if}\quad \mu_{p_j}=q \, ,\\
 p_j=1-p_j^-,&\mbox{if}\quad \mu_{p_j}=-q \, ,\\
 h_j=N'+1-h_j^+,&\mbox{if}\quad \mu_{h_j}=q \, ,\\
 h_j=h_j^-,&\mbox{if}\quad \mu_{h_j}=-q \, .
 \end{array}
 \end{equation}
All the integers  $\{p^\pm\}$ and $\{h^\pm\}$ defined above are positive and vary in a range such that
 \begin{equation}\label{domain}
 \lim_{N\to\infty}\frac{1}{N}\sum p_j^\pm=\lim_{N\to\infty}\frac{1}{N} \sum h_j^\pm =0 \, .
 \end{equation}
It means that all the particles and holes collapse to $\pm q$  in the thermodynamic limit.
Then the expression for the excitation momentum of critical states belonging to the $\mathbf{P}_\ell$
class becomes
 \begin{equation}\label{ex-Mom1}
 {\cal P}_{ex}=\frac{2\pi}{L} \ell N'
+\frac{2\pi}{L}{\cal P}_{ex}^{(d)}.
\end{equation}
Above, we have set
 \begin{equation}\label{ex-Mom-d}
 {\cal P}_{ex}^{(d)}=\sum_{j=1}^{n^+_p}p^+_j-
\sum_{j=1}^{n^-_p}p^-_j+\sum_{j=1}^{n^+_h}h^+_j-\sum_{j=1}^{n^-_h}h^-_j+n^-_p
-n^+_h.
\end{equation}
We recall  that in the thermodynamic limit $N'/L \rightarrow D$ with $D=k_{{}_F}/\pi$.

It is customary in the Bethe ansatz framework to describe excited states in terms of their shift function, see e.g. \cite{BogIK93L}. The latter characterizes
the way in which the Bethe roots (rapidities of quasi-particles) of a given excited state are shifted with respect to the ground state ones. It is in fact the result of the non trivial interactions that hold in the model, making the quasi-particles  inside the Fermi zone  move slightly due to the creation of particle/hole excitations.
In the case of a generic particle/hole excited state, this shift function depends on the rapidities of the particles $\{\mu_p\}$
and holes $\{\mu_h\}$ occurring in the given state. However, in the case of critical excited states, this shift function is  the same for all the states belonging to the same $\mathbf{P}_{\ell}$ class, and  is denoted by $F_\ell(\la)$. Of course, the function $F_\ell(\la)$ depends on the model, and on the excited state selected by the local operator under consideration. For integrable models, it is obtained as a solution of a linear integral equation\footnote{%
The linear character of the integral equation satisfied by the shift function implies that $F_\ell(\lambda)$  linearly depends on the integer $\ell$ (see section~\ref{CFSXXZ-S}).}. It was shown in \cite{Sla90,KitKMST11a} that the exponents $\theta$  can be expressed in terms of the Fermi boundary values of this function:
\[  F_{\ell}^-\equiv F_\ell(-q), \quad F_{\ell}^+\equiv F_\ell(q)+N'-N. \]
In particular the exponents $\theta$ for the form factors of the class $\mathbf{P}_\ell$ are all equal to
a same value, $\theta_\ell$, which can be written as \cite{KitKMST11a}
\begin{equation}\label{crit-theta}
\theta_\ell=(F_{\ell}^- +\ell)^2+(F_{\ell}^+ +\ell)^2.
\end{equation}
As was mentioned above the values of the shift function $F_{\ell}^\pm$
depend on the specific model and on the operators ${\cal O}_j$ that we deal with. However the functional expression  \eqref{crit-theta}
for $\theta_\ell$ in terms of the constants $F_{\ell}^\pm$ is universal and model independent.

Thus, the sum over form factors \eq{Cor-fun-ff} being restricted to the critical ones takes the form
\begin{equation}\label{Cor-fun-ell}
 \langle{\cal O}_1(x')\,{ \cal O}_2(x+x')\rangle_{cr}=\lim_{L\to\infty}\sum_{\ell=-\infty}^\infty
  L^{-\theta_\ell}\, e^{2ix\ell k_{{}_F}}\
  \hspace{-5mm}
  \sum_{\{p^\pm\},\{h^\pm\}\atop{n^+_p-n^+_h
  =n^-_h-n^-_p=\ell}}
  \hspace{-5mm}
e^{ \frac{2\pi i x}{L}{\cal P}_{ex}^{(d)}  }A^{(\ell)}(\{p^\pm\},\{h^\pm\}).
 \end{equation}
Observe that this sum is separated into two parts: the external sum over  different $\mathbf{P}_\ell$ classes,
and the sums over the quantum numbers $\{p^\pm\}$, $\{h^\pm\}$ within a fixed $\mathbf{P}_\ell$ class.
We call the last ones  {\it critical sums of order $\ell$}. Calculating these critical sums we deal only
with the discrete amplitude $A^{(\ell)}(\{p^\pm\},\{h^\pm\})$  weighted with the
exponents $\exp(\frac{2\pi i x}{L}{\cal P}_{ex}^{(d)})$. It is remarkable that $A^{(\ell)}$
has a purely kinematical interpretation and that, up to a constant factor, its functional form
is universal and model independent.

Consider the $\ell$-shifted state  $\ket{\psi'_{\ell}}$ with the Bethe roots given by the following equations:
 \begin{equation}\label{TBE-lj-ell}
 Lp_0(\mu_{j})-\sum_{k=1}^{N'}\vartheta(\mu_{j}-\mu_{k})=2\pi \left(j+\ell-
 \frac{N'+1}2\right),
 \qquad j=1,\dots,{N'}.
 \end{equation}
This excited state belongs to the $\mathbf{P}_\ell$ class.  Since all the form factors of the $\mathbf{P}_\ell$ class scale as $L^{-\theta_\ell}$, we define the renormalized amplitude $\overline{\mathcal{F}}{}^{(1)}_{{}_{\ell}}\mathcal{F}^{(2)}_{{}_{\ell}}$ which corresponds to matrix elements
of the operators ${\cal O}_1$ and ${\cal O}_2$ between the ground state and the $\ell$-shifted state $\ket{\psi'_{\ell}}$:
  \begin{equation}
\label{basic_ff}
\overline{\mathcal{F}}{}^{(1)}_{{}_{\ell}}\mathcal{F}^{(2)}_{{}_{\ell}}=\lim_{L\to\infty} L^{\theta_\ell}\frac{\bra{\psi_g}\mathcal{O}_1(x')
\ket{\psi'_{\ell}}\bra{\psi'_{\ell}}\mathcal{O}_2(x')\ket{\psi_{g}}}
{\bra{\psi_g}\psi_g\rangle\bra{\psi'_{\ell}}\psi'_{\ell}\rangle}.
%
\end{equation}
This quantity is finite in the thermodynamic limit \cite{KitKMST09c,KitKMST11a} and, due to the translation invariance, it does not depend on $x'$.

The discrete amplitudes corresponding to a given $\mathbf{P}_{\ell}$ class can be expressed in terms of the above quantity. Namely, one has \cite{KitKMST11a}
\begin{multline}
A^{(\ell)}(\{p^\pm\},\{h^\pm\}) =
 \overline{\mathcal{F}}{}^{(1)}_{{}_{\ell}}\mathcal{F}^{(2)}_{{}_{\ell}}
 \,\frac{G^2(1+F_{\ell}^+)G^2(1-F_{\ell}^-)}{G^2(1+\ell+F_{\ell}^+)G^2(1-\ell-F_{\ell}^-)}
\left(\frac{\sin(\pi F_{\ell}^+)}{\pi}\right)^{\!\!2n^+_h}\!\! \left(\frac{\sin(\pi F_{\ell}^-)}{\pi}\right)^{\!\!2n^-_h}\\
 \times
 R_{n_p^+,n_h^+}(\{p^+\},\{h^+\}|F_{\ell}^+) \;
R_{n_p^-,n_h^-}(\{p^-\},\{h^-\}|-F_{\ell}^-),\label{crit-ff-rep}
 \end{multline}
 where $G(z)$ is the Barnes function satisfying $G(z+1)=\Gamma(z)G(z)$, and
  \begin{equation}\label{def-R}
 R_{n,n'}(\{p\},\{h\}|F)= \frac{\prod\limits_{j>k}^n(p_j-p_k)^2\prod\limits_{j>k}^{n'}(h_j-h_k)^2}
 {\prod\limits_{j=1}^n\prod\limits_{k=1}^{n'}(p_j+h_k-1)^2} \;
 \prod_{k=1}^{n}\frac{\Gamma^2(p_k+F)}{\Gamma^2(p_k)}
 \prod_{k=1}^{n'}\frac{\Gamma^2(h_k-F)}{\Gamma^2(h_k)}.
  \end{equation}
Let us comment the formulas above.

The representation \eqref{crit-ff-rep} was derived straightforwardly for the particle-hole form factors in the $XXZ$ Heisenberg chain \cite{KitKMST11a} and quantum one-dimensional Bose gas \cite{KozT11}. It represents the thermodynamic limit of known determinant formulas for form factors for finite $N$ and $L$  \cite{KojKS1997,KitMT99}.
In this representation, only the constant $\overline{\mathcal{F}}{}^{(1)}_{{}_{\ell}}\mathcal{F}^{(2)}_{{}_{\ell}}$   depends on the model and on the operators considered. The rest of the equation \eqref{crit-ff-rep} is universal.
This occurs because the particle--hole form factors for both models are proportional to generalized Cauchy determinants, namely,
 \begin{equation}\label{Cauchy}
 \mathcal{F}_{\psi_g \psi'}\sim \frac{\prod\limits_{j>k}^{N}(\lambda_j-\lambda_k)
 \prod\limits_{j>k}^{N'}(\mu_{\ell_j}-\mu_{\ell_k})}
 {\prod\limits_{j=1}^N\prod\limits_{k=1}^{N'}(\lambda_j-\mu_{\ell_k})}.
 \end{equation}
We call the {\it rhs} of \eqref{Cauchy} the generalized Cauchy determinant because for $N'=N$ it
gives known expression for the standard Cauchy determinant $\det[1/(\lambda_j-\mu_{\ell_k})]$ in terms of the rapidities $\lambda_j$ of the quasi-particles in the ground state and $\mu_{\ell_k}$ in the excited state.
The appearance of such an object  in the formulas for form factors has a clear
physical interpretation. The fact that $\mathcal{F}_{\psi_g \psi'}=0$ as soon as $\lambda_j=\lambda_k$ or
$\mu_{\ell_j}=\mu_{\ell_k}$ simply takes into account the fermionic structure of the particle/hole description of the ground state and of the excitations. The singularities at $\lambda_j=\mu_{\ell_k}$ produce, in the thermodynamic limit,  poles occurring whenever the rapidities of particles and holes coincide at the Fermi boundaries.
Such a behavior near the Fermi boundary has a purely kinematical origin and is quite universal. It  appears for example in the form factor bootstrap program. There the residues of the $n$-particle form factor  at the particle/antiparticle poles are given in terms of $(n-1)$-particle form factors \cite{Smi92b}.

On the other hand it is not difficult to check (see e.g. \cite{KitKMST11a}) that the combination of the generalized
Cauchy determinants corresponding to the  ratios $({\cal F}^{(1)}_{\psi_g\,\psi'}{\cal F}^{(2)}_{\psi'\,\psi_g} ) /(\overline{\mathcal{F}}{}^{(1)}_{{}_{\ell}}\mathcal{F}^{(2)}_{{}_{\ell}})$ of form factors gives, in the thermodynamic limit, exactly
the universal part of the amplitude  \eqref{crit-ff-rep}, provided ${\cal F}^{(1)}_{\psi_g\,\psi'}$ and ${\cal F}^{(2)}_{\psi'\,\psi_g}$ are critical form factors of the $\mathbf{P}_\ell$ class. Thus, a rather complicated, at first sight, dependence of the form factors on the quantum numbers such as in \eqref{crit-ff-rep} has a simple physical origin.

One can observe that the expression \eqref{crit-ff-rep} is factorized into the product of two parts corresponding to the left and right Fermi boundaries. Such factorization occurs in the thermodynamic limit, when we neglect all
corrections of order $o(1)$ at $L\to\infty$. If we do not neglect these corrections,  then we evidently produce
subleading contributions to the form factors of order $L^{-\theta_\ell-1}$, $L^{-\theta_\ell-2}$ etc. On the
other hand in this case we do not have a complete factorization of the final answer, like in \eqref{crit-ff-rep}, and thus, we can say that higher order corrections describe an effective coupling between the two Fermi boundaries.

Using the representation \eqref{crit-ff-rep} we can present the sum of the critical form factors in the following form:
\begin{multline}
 \langle{\cal O}_1(x')\,{ \cal O}_2(x+x')\rangle_{cr}=\lim_{L\to\infty}\sum_{\ell=-\infty}^\infty
  L^{-\theta_\ell }\, e^{2ix\ell k_{{}_F}}
  \overline{\mathcal{F}}{}^{(1)}_{{}_{\ell}}\mathcal{F}^{(2)}_{{}_{\ell}}
  \,\frac{G^2(1+F_{\ell}^+)G^2(1-F_{\ell}^-)}{G^2(1+\ell+F_{\ell}^+)G^2(1-\ell-F_{\ell}^-)}\\
  \times
   \sum_{\{p^\pm\},\{h^\pm\}\atop{n^+_p-n^+_h=n^-_h-n^-_p=\ell}}
   \hspace{-2mm}
   \exp\left[ \frac{2\pi i x}{L}{\cal P}_{ex}^{(d)}  \right]\,\left(\frac{\sin(\pi F_{\ell}^+)}{\pi}\right)^{\! 2n^+_h} \left(\frac{\sin(\pi F_{\ell}^-)}{\pi}\right)^{\! 2n^-_h} \\
 \times
 R_{n_p^+,n_h^+}(\{p^+\},\{h^+\}|F_{\ell}^+) \;
R_{n_p^-,n_h^-}(\{p^-\},\{h^-\}|-F_{\ell}^-).
\label{Cor-fun-ell_F}
 \end{multline}

We would like to mention that this equation gives explicitly the contribution of the critical form factors in the thermodynamic limit.

It is important to notice that the critical sums like in \eqref{Cor-fun-ell_F} are universal and hence appear already in the models equivalent to free fermions. There, of course, the shift function is a trivial constant. Indeed, in free models one cannot obtain a non-trivial shift function because of the absence of interaction.  Therefore in such models a shift between the ground state quasi-particles and the ones describing the excited state either occurs due to non-vanishing difference $N-N'$, or it can be created
artificially by considering {\it twisted excited states} (see section~\ref{CFSXXZ-S}). In contrast, for the interacting
models we deal with, a non-trivial shift function arises automatically.  It is remarkable, however, that being restricted to the states belonging to a fixed $\mathbf{P}_\ell$ class, this non-trivial shift function enters the critical sums only through its values at the Fermi boundaries, namely through the two constants
$F_{\ell}^\pm$. In this sense one can say that we reduced
our model to an effective (deformed) free fermion one in every $\mathbf{P}_\ell$ class. We stress however, that for interacting
systems such a reduction cannot be carried out uniformly for all critical form factors, in contrast to what happens for a free theory: every $\mathbf{P}_\ell$ class
of form factors should be described by its own (deformed by the values $F_{\ell}^\pm$) free fermion theory.

The most important property of the critical sums is that they can be explicitly computed due to the following summation formula:
\begin{multline}
 \sum_{n,n'\ge 0\atop{n-n'=\ell}}\,
 \sum_{ \substack{ p_1<\cdots<p_{n} \\ p_a \in \mathbb{N^*}}  }\,
 \sum_{ \substack{ h_1<\cdots<h_{n'} \\ h_a \in \mathbb{N^*} } }
 \exp\!\! \bigg[ \frac{2\pi i x}{L}\bigg(\sul_{j=1}^{n}(p_j-1)+\sul_{k=1}^{n'} h_k\bigg)\bigg]
 \left(\frac{\sin\pi F}\pi\right)^{2n'} R_{n,n'}(\{p\},\{h\}|F) \\
=\frac{G^2(1+\ell+F)}{G^2(1+F)}\,\frac{\exp\Big[\frac{\pi i x}{L}\ell(\ell-1)\Big]\,\,}
{\Big(1-\exp\left[ \frac{2\pi i x}{L}\right]\Big)^{(F+\ell)^2}}.
\label{magic formula}
 \end{multline}
We prove \eqref{magic formula} in  Appendix A. Amazingly, for the case $\ell=0$, such a summation formula
appeared already in the problem  of defining $Z$-measures on partitions \cite{KerOV93,Ker2000,BorO2000,BorO2000a,Oko01,BorO2001,Ols2003}\footnote{We are grateful to O. Lisovyy for pointing out these works to us.}. Note that this formula was also  useful in the case of thermal correlation functions of the Bose gas \cite{KozMS11b}.

Using this identity we obtain for the two-point function
  \begin{align}
 \langle{\cal O}_1(x')\,{ \cal O}_2(x+x')\rangle_{cr}=
 \lim_{L\to\infty}\sum_{\ell=-\infty}^\infty
 \frac{L^{-\theta_\ell }\; e^{2ix\ell k_{{}_F}}\overline{\mathcal{F}}{}^{(1)}_{{}_{\ell}}\mathcal{F}^{(2)}_{{}_{\ell}}  }
{ \Big(1-e^{ \frac{2\pi i x}{L} }\Big)^{(F_{\ell}^++\ell)^2}
\Big(1-e^{ - \frac{2\pi i x}{L} }\Big)^{(F_{\ell}^-+\ell)^2}  }  Ê .
%
%
\label{Cor-fun-ell_MF}
 \end{align}
Now the thermodynamic limit can be easily taken using \eqref{crit-theta} and we obtain our final result:
   \begin{equation}
 \langle{\cal O}_1(x')\,{ \cal O}_2(x+x')\rangle_{cr}=\sum_{\ell=-\infty}^\infty
 e^{2ix\ell k_{{}_F}+i \frac\pi 2 \mathrm{sign}(x) [(F_{\ell}^-+\ell)^2-(F_{\ell}^++\ell)^2]}
  \frac{\overline{\mathcal{F}}{}^{(1)}_{{}_{\ell}}\mathcal{F}^{(2)}_{{}_{\ell}}}{\big(2\pi  |x|\big)^{\theta_\ell}}.
\label{Cor-fun-ell_MF_As}
 \end{equation}

Observe that the critical exponents $\theta_\ell$ driving the leading power law decay in
\eqref{Cor-fun-ell_MF_As} coincide with the ones describing the scaling behavior of the
form factors (in terms of the system size). This property is quite natural as the distance $x$ and
the system size $L$ in \eqref{magic formula} appear only in the combination $x/L$. As soon as we expect
the sum over critical form factors to compensate the vanishing coefficient $L^{-\theta_\ell}$,
it should produce a factor $(x/L)^{-\theta_\ell}$, leading to an answer like in \eqref{Cor-fun-ell_MF_As}.
The $1/L$-corrections to the form factors describing the coupling of the  two Fermi boundaries,
and mentioned above, should produce respective $1/x$-corrections to the leading asymptotic behavior of each oscillating harmonic in \eqref{Cor-fun-ell_MF_As}. In the CFT language such higher-order corrections
should correspond to the contribution of the descendants of primary fields.

We would like to underline once again that this is an explicit result for the contribution of the critical form factors to the two-point
functions of a very large class of integrable critical models.
Our main conjecture states  that this equation gives the leading asymptotic behavior of all the
oscillation harmonics of the two-point function.  This conjecture is supported by the fact that, in the case of the $XXZ$ chain (or the
quantum non-linear Schr\"{o}dinger model), we reproduce the exact answer stemming from a Riemann--Hilbert problem analysis
for the first few harmonics. Also, our asymptotic formula reproduces the CFT predictions for the $XXZ$ spin chain and
quantum non-linear Schr\"{o}dinger model
for all the oscillating harmonics.

Our method  gives a simple explanation of the main result of \cite{KitKMST09b,KitKMST09c}. We have shown there that the amplitude of the leading oscillating term of the spin--spin  correlation function for the $XXZ$ spin chain (or of the density-density correlation  function for the quantum non-linear Schr\"odinger model) can be expressed in terms of a rescaled special form factor. Moreover we observed that the critical exponent driving this leading oscillating term is equal to the one used for the rescaling of this special form factor.  The critical form factor approach gives a simple explanation for both results, generalizing it to the leading asymptotic behavior of all the oscillating harmonics. Recently another explanation of these facts was given in \cite{ShaGCI11}.

Let us conclude this section by one remark about the non-oscillating ($\ell=0$) term in \eqref{Cor-fun-ell_MF_As}. For some correlation functions the
corresponding exponent is zero $\theta_0=0$; thus  the corresponding term is constant. A small modification of the critical form factor
approach  however permits us to compute the first non-oscillating correction to this formula. It requires the notion of a `twisted propagator'
and `twisted Bethe states'. In section~\ref{CFSXXZ-S} we will show how these objects can be introduced for the $XXZ$ spin chain.


\section{Correlation functions of spins in the $XXZ$ chain\label{CFSXXZ-S}}

In this section, we apply the general method described above to the
spin--spin correlation functions of the $XXZ$ spin-$1/2$ chain.  Recall that this model is given by
the Hamiltonian
 \begin{equation}\label{0-HamXXZ}
 H=\sum_{k=1}^{L}\left(
 \sigma^x_{k}\sigma^x_{k+1}+\sigma^y_{k}\sigma^y_{k+1}
 +\Delta(\sigma^z_{k}\sigma^z_{k+1}-1)\right)-\frac{h}{2}\sum_{k=1}^{L}\sigma^z_{k}.
 \end{equation}
Here $\sigma^{x,y,z}_{k}$ are the spin operators (Pauli matrices) acting on the $k$-th site
of the chain,  $h$ is an external magnetic field and the model is subject to periodic boundary conditions.
The $XXZ$ spin chain above exhibits different phases depending on the value of  the anisotropy parameter  $\Delta$.
Here, we only focus on the critical regime $-1<\De<1$ where we set $\Delta=\cos \zeta$.

In the following, we consider the two correlation functions\footnote{$\langle\sigma^-_1\sigma^+_{m+1}\rangle$ can be obtained from $\langle\sigma^+_1\sigma^-_{m+1}\rangle$ by changing $m$ into $-m$.} $\langle\sigma^z_1\sigma^z_{m+1}\rangle$ and
$\langle\sigma^+_1\sigma^-_{m+1}\rangle$, with $\sigma^\pm=\frac12(\sigma^x\pm i\sigma^y)$.
Note that, since the lattice spacing is discrete, we
have denoted it by $m$ (instead of $x$ as in the previous sections).

\subsection{The correlation function $\langle \sigma_{1}^+\sigma_{m+1}^{-}\rangle$  \label{sigma-pm}}

The correlation function $\langle \sigma_{1}^+\sigma_{m+1}^{-}\rangle$ can be computed using the general method.

The only excited states which produce non-zero contributions here are the Bethe states with $N'=N+1$.
The critical values of the shift function were computed in \cite{KitKMST11a}:
\begin{equation}
F_{\ell}^-=\ell(\cZ-1)-\frac 1 {2\mathcal{Z}},\qquad F_{\ell}^+=\ell(\cZ-1)+\frac 1 {2\mathcal{Z}},
\end{equation}
where ${\cal Z}=Z(\pm q)$ is the Fermi boundary value of the dressed charge  $Z(\la)$ given by the following integral equation:
\begin{equation}
Z(\la)+\frac 1{2\pi}\int\limits_{-q}^q d\mu \,\frac {\sin2\zeta}{\sinh(\la-\mu+i\zeta)\sinh(\la-\mu-i\zeta)}\,Z(\mu)=1.
\end{equation}
The critical exponents are  given by
\begin{equation}
\theta_\ell=2\ell^2\cZ^2+\frac 1 {2\mathcal{Z}^2}.
\end{equation}
We introduce the simplest form factor of the $\mathbf{P}_\ell$ class:
  \begin{equation}
\label{basic_ff_sigma_+}
\left|\mathcal{F}^+_{{}_{\ell}}\right|^2=\lim_{L\to\infty} L^{\left(2\ell^2\mathcal{Z}^2+\frac 1 {2\mathcal{Z}^2}\right)}\frac{\left|\bra{\psi_g}\sigma^+_1|\psi'_{\ell}\rangle\right|^2}{\bra{\psi_g}\psi_g\rangle\bra{\psi'_{\ell}}\psi'_{\ell}\rangle}.
\end{equation}
The rapidities corresponding to the $\ell$-shifted state solve the following logarithmic Bethe equations:
 \begin{equation}\label{TBE-lj-ell-sigma+}
 Lp_0(\mu_{j})-\sum_{k=1}^{N+1}\vartheta(\mu_{j}-\mu_{k})=2\pi \left(j+\ell+1-
 \frac{N}2\right),
 \qquad j=1,\dots,{N+1}.
 \end{equation}
 The bare momentum and the bare scattering phase for the $XXZ$ model are
  \begin{equation}\label{1-p0}
 p_0(\lambda)=i\log\left(\frac{\sinh(\textstyle{\frac{i\zeta}2}+\lambda)}
 {\sinh(\textstyle{\frac{i\zeta}2}-\lambda)}\right),\qquad
 \vartheta(\lambda)=i\log\left(\frac{\sinh(i\zeta+\lambda)}
 {\sinh(i\zeta-\lambda)}\right).
 \end{equation}

Applying the general formula \eqref{Cor-fun-ell_MF_As}, we obtain the asymptotic result\footnote{%
There is an extra general $(-1)^m$ factor, which can be removed by
the re-definition of the Hamiltonian.}
 \begin{equation}
\langle \sigma_{1}^+\sigma_{m+1}^{-}\rangle_{cr}=\frac{(-1)^m}{\big(2\pi  m\big)^{\frac 1 {2\mathcal{Z}^2}}}\sum_{\ell=-\infty}^\infty (-1)^\ell
 e^{2im\ell\, k_{{}_F}}\,\left|\mathcal{F}^+_{{}_{\ell}}\right|^2\,\frac{1}{\big(2\pi  m\big)^{2\ell^2\mathcal{Z}^2}}.
\end{equation}
The dominant term of this expansion ($\ell=0$)  behaves as predicted by the Luttinger liquid approach  and CFT.

\subsection{The correlation function $\langle\sigma^z_1\sigma^z_{m+1}\rangle$}

In order to compute the two-point function $\langle\sigma^z_1\sigma^z_{m+1}\rangle$, it is convenient to
introduce its generating function $\langle e^{2\pi i\alpha{\cal Q}_m}\rangle$ \cite{IzeK85,ColIKT93,KitMST02a,KitMST05a}, where
 \begin{equation}\label{def-Q}
{\cal Q}_m=\frac12\sum_{k=1}^m(1-\sigma^z_k).
\end{equation}
This function generates the two-point function $\langle\sigma^z_1\sigma^z_{m+1}\rangle$ via
 \begin{equation}\label{ss-ebQ}
\langle\sigma^z_1\sigma^z_{m+1}\rangle=-\frac1{2\pi^2}\left.
\partial^2_\alpha  \mathbf{D}^2_m \langle e^{2\pi i\alpha{\cal Q}_m}\rangle\right|_{\alpha=0}-2D+1,
\end{equation}
where $\mathbf{ D}^2_m$ is the second lattice derivative.

The large $m$  behavior of $\langle e^{2\pi i\alpha{\cal Q}_m}\rangle$ can be obtained in the framework of the form
factor approach. For this one introduces the complete set of eigenstates of the {\it twisted transfer matrix} ${\cal
T}_\kappa(\lambda)$ (see e.g. \cite{KitMST05a}), where $\kappa=e^{2\pi i\alpha}$ is a
complex (twist) parameter. This transfer matrix produces an $XXZ$ spin-$1/2$ Hamiltonian but subject to quasi-periodic (instead of periodic)
boundary conditions. Hereby the excited states of the Hamiltonian \eqref{0-HamXXZ} are obtained as the $\alpha\to0$ limit
of the eigenstates of the twisted transfer matrix  ${\cal
T}_\kappa(\lambda)$ \cite{TarV95,KitMST05a,KitKMST09c}.
The introduction of the extra parameter $\kappa$ is very useful in the analysis of the form factor series.

In complete analogy with the standard algebraic Bethe ansatz
considerations, the eigenstates $|\psi_\kappa(\{\mu\})\rangle$ of
${\cal T}_\kappa(\lambda)$ can be parameterized by sets of solutions
of the twisted Bethe equations
 \begin{equation}\label{TBE-lj-alpha}
 Lp_0(\mu_{j})-\sum_{k=1}^{N}\vartheta(\mu_{j}-\mu_{k})=2\pi \left(n_j+\alpha-
 \frac{N+1}2\right),
 \qquad j=1,\dots,{N}.
 \end{equation}
For this correlation function only the form factors with $N'=N$ produce non-zero contributions.

From the solution of the quantum inverse problem it can be easily shown that
for any eigenstate of the twisted transfer matrix the form factors are proportional to the scalar products
\begin{equation}
\bra{\psi_\kappa(\{\mu\})}e^{2\pi i\alpha{\cal Q}_m}|\psi_g\rangle=e^{i\mathcal{P}_{ex}m}\bra{\psi_\kappa(\{\mu\})}\psi_g\rangle,
\end{equation}
where ${\cal P}_{ex}$ is the momentum of the excited state.

Critical form factors related to the operator $e^{2\pi i\alpha{\cal Q}_m}$ were computed in
\cite{KitKMST11a}.  For the $\mathbf{P}_\ell$ class, the boundary values of shift function $F_\ell$ are
\begin{equation}
F_{\ell}^-=F_{\ell}^+=\a \mathcal{Z}+\ell(\mathcal{Z}-1).
\end{equation}
Thus the exponent $\theta_{\ell}(\alpha)$ can be written as
 \begin{equation}\label{theta-zz}
 \theta_{\ell}(\alpha)=2\Big((\alpha+\ell){\cal Z}\Big)^2.
\end{equation}
The amplitudes in this case are just the  normalized scalar products
 \begin{equation}
\label{basic_ff_alphaQ}
\left|\mathcal{F}_{{}_{\a+\ell}}\right|^2=\lim_{L\to\infty} L^{\theta_{\ell}(\alpha)}\frac{\left|\bra{\psi_g}\psi'_{\a+\ell}\rangle\right|^2}{\bra{\psi_g}\psi_g\rangle\bra{\psi'_{\a+\ell}}\psi'_{\a+\ell}\rangle},
\end{equation}
where $\ket{\psi'_{\a+\ell}}$ is the state parameterized by the Bethe roots of
\begin{equation}\label{TBE-lj-alpha-ell}
 Lp_0(\mu_{j})-\sum_{k=1}^{N}\vartheta(\mu_{j}-\mu_{k})=2\pi \left(j+\alpha+\ell-
 \frac{N+1}2\right),
 \qquad j=1,\dots,{N}.
 \end{equation}
Note that $\bra{\psi_g}\psi'_{\a+\ell}\rangle=0$ at $\alpha=0$ and $\ell\ne 0$, as the scalar product of two different
eigenstates.

Now,  using  the general result \eqref{Cor-fun-ell_MF_As}, we obtain the leading asymptotic terms for all the oscillating harmonics:
\begin{equation}\label{asy-eaQ}
\langle e^{2\pi i\alpha{\cal Q}_m}\rangle_{cr}=\sum_{\ell=-\infty}^\infty
 e^{2im(\alpha+\ell) k_{{}_F}}
\left|\mathcal{F}_{{}_{\a+\ell}}\right|^2\,\frac{1}{(2\pi  m)^{\theta_{\ell}(\alpha)}}.
\end{equation}
The dominant terms with $\ell=0,\pm1$ were already obtained by the Riemann--Hilbert analysis
of the multiple-integral representations \cite{KitKMST09b,KitKMST09c}.
Here, we get in a much simpler way the same result as well as the natural extension to other harmonics reconstructing the $\a$-periodicity of the correlation function.

To obtain an asymptotic expansion for the two-point function we should take the $\alpha$-de\-ri\-va\-tives and lattice derivatives of \eqref{asy-eaQ}. Here one should remember that the coefficients $|\mathcal{F}_{{}_{\a+\ell}}|^2$ with $\ell\ne 0$
have a second-order zero at $\alpha=0$, while the coefficient $|\mathcal{F}_{{}_{\a}}|^2$ does not vanish in this point. This leads to the  following result:
 \begin{equation}
 \langle\sigma_1^z\sigma_{m+1}^z\rangle_{cr}=(2D-1)^2-
 \frac{2{\cal Z}^2}{\pi^2 { m}^2}+2\sul_{\ell=1}^{\infty}\cos (2m\ell k_{{}_F})\left|\mathcal{F}^z_{{}_{\ell}}\right|^2\,\frac{1}{\big(2\pi  m\big)^{2\ell^2\mathcal{Z}^2}},
\end{equation}
where
 \begin{equation}
\label{basic_ff_sigma_z}
\left|\mathcal{F}^z_{{}_{\ell}}\right|^2=\lim_{L\to\infty} L^{2\ell^2\mathcal{Z}^2}\frac{\left|\bra{\psi_g}\sigma^z_1|\psi'_{\ell}\rangle\right|^2}
{\bra{\psi_g}\psi_g\rangle\bra{\psi'_{\ell}}\psi'_{\ell}\rangle},
\end{equation}
and the Bethe roots corresponding to the $\ell$-shifted state are defined by \eqref{TBE-lj-alpha-ell} with $\alpha=0$.

The first terms of this expansion (non-oscillating terms and $\ell=1$ term) reproduce the asymptotic behavior predicted by various methods (Luttinger liquid approach \cite{LutP75,Hal80,Hal81a,Hal81b}, CFT \cite{BelPZ84,Aff85,BloCN86,Car84,Car86}). The amplitude of the first oscillating term is given by the special form factor as predicted by the Riemann--Hilbert approach \cite{KitKMST09b,KitKMST09c}.

We would like to underline that from the beginning we could apply the general method directly to the two-point function without considering the generating function $\langle e^{2\pi i\alpha{\cal Q}_m}\rangle$. However the method proposed here
(the use of twisted eigenstates) has one evident advantage: it produces the first subdominant  non-oscillating term of order $\frac 1{m^2}$ (the dominant non-oscillating term being a constant).

\section*{Conclusion}

In this paper we have developed a new method for computing  the contribution of the critical form factors to the asymptotic behavior of  two-point correlation functions of  critical integrable models. This result is quite general and can be applied to a very large class of integrable field theories and lattice models. We have shown that for the Heisenberg spin chain this contribution reproduces the CFT predictions for the asymptotic behavior of the correlation functions. Exactly the same result can be obtained for the quantum non-linear  Schr\"odinger model,  reproducing once again existing predictions for the asymptotic behavior.

We hope that the method that we presented here can be applied to a wide class of non-integrable systems as well. We have pointed out already that, within every fixed $\mathbf{P}_\ell$ class,  the critical sum over form factors coincides with the one corresponding to an effective (deformed)
free fermion model. Therefore, if we have a general quantum system which admits, for every fixed excitation momentum, a (deformed) free fermion interpretation of its low energy spectrum, we can expect  the sum over form factors within this sector of excitations to coincide with the critical sums considered in this paper. Then, calculating these sums explicitly via \eqref{magic formula}, one obtains the leading critical exponents for all oscillating harmonics in the asymptotic expansion of the correlation function.

We would like to stress once more that  effective free models describing the critical form factors of interacting systems are different for different $\mathbf{P}_\ell$ classes. In particular, the values of the shift function on the Fermi boundaries $F_{\ell}^\pm$ change from one class to another. Nevertheless, these values are certain constants for all excited states of the same class. It is quite natural to expect that, in the thermodynamic limit,  these constants will only depend on the $\mathbf{P}_\ell$ class of the excited states, and not on other details, in much the same way as in the integrable situation. In fact, the shift function just encodes the polarization of the Dirac sea right at the Fermi boundaries as the result of the microscopic interactions at work in the system considered. Then all the machinery that we described for the integrable situation will lead to the computation of the asymptotic behavior of the correlation function. The only difference comes
from the fact that, in that case, the shift function $F$ does not satisfy some exact linear integral equation and that its value at the Fermi boundaries could be hard to compute. However this quantity should be just considered as a physical parameter to be plugged into the summation formula \eqref{magic formula} to  determine the corresponding asymptotic behavior of the correlation function. This picture gives a nice (fully interacting) explanation of the origin of the critical exponents: they are just the result of the polarization of the Dirac sea  at the Fermi boundaries produced by the creation of particle--hole excitations. Moreover, there is here some notion of universality. Generically different models possess different shift functions. However, models having the same values $F_{\ell}^\pm$ will have the same critical behavior.
We believe that, all together, this scheme provides a microscopic particle-type description of  CFT starting from lattice models that should be useful both from conceptual and practical viewpoints.

In a forthcoming publication we will show how this method can be extended to study the time dependent correlation functions.


\section*{Acknowledgements}

J.M.M., N.S. and V.T. are supported by CNRS.
N.K, J.M.M. and V.T. are supported by ANR grant  ANR-10-BLAN-0120-04-DIADEMS.
N.K. is supported by the Burgundy region, FABER grant 2010-9201AAO047S00753.
We also acknowledge  the support from the GDRI-471 of CNRS ``French-Russian network in Theoretical and Mathematical  Physics" and RFBR-CNRS-09-01-93106L-a.
N.S. is also supported by the Program of RAS Mathematical Methods of the Nonlinear Dynamics,
RFBR-11-01-00440, SS-8265.2010.1.
When this work was done, K.K.K. was supported by the EU Marie-Curie Excellence Grant MEXT-CT-2006-042695 and DESY.
N.K., N.S. and K.K.K. would like to thank the Theoretical Physics group of the Laboratory of Physics at ENS Lyon for hospitality, which makes this collaboration possible.
N.K., J.M.M. and V.T. would  like to thank LPTHE (Paris VI University) for hospitality.

\appendix
\section{Proof of the summation formula}
\label{Appendix proof summation formula}

In this appendix we give the proof of the formula for the critical sums \eqref{magic formula}. We start with the simplest and most fundamental case $\ell=0$. Given $|w|<1$ and a complex valued $\nu$, we define the function $f_0(\nu,w)$ by the series
\begin{multline}\label{Ext-For}
f_0(\nu,w)=\sum_{n=0}^\infty\quad
\sum_{ \substack{ p_1<\cdots<p_n \\ p_a \in \mathbb{N}^*}  }\quad
\sum_{ \substack{  h_1<\cdots<h_n \\ h_a \in \mathbb{N}^* } }
w^{\sum_{j=1}^n(p_j+h_j-1)}\left(\frac{\sin\pi
\nu}\pi\right)^{2n}\\
\times\left(\det_n\frac1{p_j+h_k-1}\right)^2\prod_{k=1}^{n}
\frac{\Gamma^2(p_k+\nu)\Gamma^2(h_k-\nu)}{\Gamma^2(p_k)\Gamma^2(h_k)}.
\end{multline}
The summand is a symmetric function of the variables $\{p\}$ and likewise a symmetric function
of the variables $\{h\}$. Moreover it is vanishing whenever two of the  $p$-type variables or two of the $h$-type variables
coincide. Therefore,
\begin{equation}\label{repl-1}
\sum_{\substack{ p_1<\cdots<p_n \\ p_a\in \mathbb{N}^* }  }\;
\sum_{\substack{ h_1<\cdots<h_n \\ h_a\in \mathbb{N}^* } }=
\frac1{(n!)^2}\sum_{  p_1,\dots,p_n  =1   }^{\infty}
\;\sum_{  h_1,\dots,h_n =1  }^{\infty}.
 \end{equation}
Notice that, given any symmetric function $ {\cal G}(\{p\},\{h\})$ of the $\{p\}$ and  $\{h\}$ variables, one has
 \begin{multline}\label{repl-2}
 \sum_{p_1,\dots,p_n=1}^{\infty}\;\sum_{
 h_1,\dots,h_n=1}^{\infty}\left(\det_n\frac1{p_j+h_k-1}\right)^2\cdot
 {\cal G}(\{p\},\{h\})\\
 = n!\sum_{p_1,\dots,p_n=1}^{\infty}\;\sum_{
 h_1,\dots,h_n=1}^{\infty}\left(\det_n\frac1{p_j+h_k-1}\right)\prod_{j=1}^n\frac1{p_j+h_j-1}
 \cdot {\cal G}(\{p\},\{h\}) .
 \end{multline}
This property allows one to recast the series \eqref{Ext-For} as
\begin{equation}\label{sum-ov-h}
 f_0(\nu,w)=\sum_{n=0}^{\infty}\frac1{n!}\sum_{
 h_1,\dots,h_n=1}^{\infty}\det_{j=1,\dots,n\atop{k=1,\dots,n}} V(h_j,h_k),
 \end{equation}
where
\begin{equation}\label{Vjk}
 V(h_j,h_k)=\sum_{p=0}^\infty\left(\frac{w^{p+(h_k+h_j)/2}}{(p+h_k)(p+h_j)}\frac{\Gamma^2(p+1+\nu)}{\Gamma^2(p+1)}\right)
 %
 \left(\frac{\sin\pi
 \nu}\pi\right)^{2}
 \frac{\Gamma(h_k-\nu)\Gamma(h_j-\nu)}{\Gamma(h_k)\Gamma(h_j)}.
\end{equation}
 By definition,  (\ref{sum-ov-h}) gives the determinant of the following infinite matrix:
 \begin{equation}\label{inf-det}
 f_0(\nu,w)=\det_{j=1,\dots,\infty\atop{k=1,\dots,\infty}}[\delta_{jk}+
 V(j,k)].
 \end{equation}

This determinant can be computed for general $\nu$ and $w$.
In fact, whenever $\nu$ is integer, this infinite determinant boils down to a finite size determinant
which can be computed using the Cauchy determinant formula.
In the general case, however, no such simplification arises and one has to build the
computation on  known determinant evaluations of Toeplitz and Hankel operators obtained by H. Widom \cite{Wid75}.
Below we consider the two cases separately.

 \subsection{Integer $\nu$}

Making the replacement $h_j\leftrightarrow p_j$, we conclude that
$f_0(\nu,w)=f_0(-\nu,w)$. Therefore it is sufficient to consider the
case $\nu=N$, where $N\in\mathbb{Z}_+$.

If $\nu=N$, then the combination $\sin^{2}(\pi \nu)\,
\Gamma(h_k-\nu)\Gamma(h_j-\nu)$ vanishes as soon as $h_j>N$ or
$h_k>N$. Thus, the series in each $h_j$ in \eqref{sum-ov-h} becomes truncated
at $h_j=N$, and we obtain the determinant of an
$N\times N$ matrix:
\begin{equation}\label{NN-det}
 f_0(\nu,w)=\det_{j=1,\dots,N\atop{k=1,\dots,N}}\left[\delta_{jk}+
 V(j,k)\right].
 \end{equation}
 Setting $\nu=N$ in \eqref{Vjk} we get, after simple algebra,
\begin{equation}\label{Vjk-simp}
 V(j,k)=\frac{w^{(j+k)/2}}{\prod\limits_{m=1\atop{m\ne
j}}^N(j-m)\prod\limits_{m=1\atop{m\ne k}}^N(k-m)}
\sum_{p=0}^\infty w^{p} \frac{\prod\limits_{m=1}^N(p+m)^2}{(p+k)(p+j)}.
\end{equation}

Let us transform the $\det(I+V)$ obtained as
 \begin{equation}\label{AMB}
 \det(I+V)=\frac{\det_N(AA^T+AVA^T)}{(\det_N A)^2},
 \end{equation}
with
 \begin{equation}\label{AB}
 A_{jk}=w^{-k/2} k^{j-1},\qquad j,k=1,\dots,N.
\end{equation}
 Obviously,
 \begin{equation}\label{detAB}
 (\det_N A)^2=w^{-N(N+1)/2}\prod_{j>k}^N(j-k)^2=w^{-N(N+1)/2}\prod_{k=0}^{N-1}(k!)^2.
 \end{equation}
To compute the product $AVA^T$ one can use the following identity:
\begin{equation}\label{ident}
\sum_{\ell=1}^N\frac{\ell^{s-1}}{(\ell+p)\prod\limits_{\substack{m=1\\ m\ne
\ell}}^N(\ell-m)}=(-1)^{N+s}\frac{p^{s-1}}{\prod\limits_{m=1}^N(p+m)},\qquad
s=1,\dots,N,
\end{equation}
which follows from
 \begin{equation}\label{ident-int}
 \oint\limits_{|z|=R}\frac{z^{s-1}\,dz}{(z+p)\prod\limits_{m=1}^N(z-m)}=0,\qquad
 s=1,\dots,N, \quad \text{with} \;\; R> \max( N,p) .
\end{equation}
Then we obtain
\begin{equation}\label{new-det}
f_0(\nu,w)=\frac{w^{N(N+1)/2}}{\prod\limits_{k=0}^{N-1}(k!)^2}
\det_N\bigg[ \sum_{p=1}^Nw^{-p}p^{j+k-2}+(-1)^{j+k}\sum_{p=0}^\infty w^{p}p^{j+k-2}\bigg].
\end{equation}
Setting here $w=e^{-t}$, we get
\begin{multline}\label{new-det1}
f_0(\nu,w)=\frac{e^{-tN(N+1)/2}}{\prod\limits_{k=0}^{N-1}(k!)^2}
\det_N\bigg[\partial_t^{j+k-2}\Big(\sum_{p=1}^Ne^{pt}+\sum_{p=0}^\infty e^{-pt}\Big)\bigg]\\
=\frac{e^{-tN(N+1)/2}}{\prod\limits_{k=0}^{N-1}(k!)^2}
\det_N\bigg[ \partial_t^{j+k-2}\frac{e^{Nt}}{1-e^{-t}}\bigg] .
\end{multline}

We now use that
\begin{equation}\label{hom-limF}
 \lim_{u_1,\dots,u_N\to u_0\atop{v_1,\dots,v_N\to
 v_0}}\frac{\det_N\Phi(u_j,v_k)}{\Delta(u)
 \Delta(v)}=\prod\limits_{k=0}^{N-1}\frac1{(k!)^{2}}
 \det_N\bigg[ \frac{\partial^{j+k-2} \Phi(u_0,v_0)}{\partial u^{j-1}_0\partial v^{k-1}_0} \bigg],
\end{equation}
where $\Phi(u,v)$ is any two-variable function such that
all derivatives entering \eqref{hom-limF} exist. The notation
$\Delta$ means the Vandermonde determinants of the corresponding
variables.

Clearly the equation \eqref{new-det1} can be written as  a
homogeneous limit of the type \eqref{hom-limF}:
\begin{equation}\label{hom-lim}
f_0(\nu,e^{-t})=e^{-tN(N+1)/2}\lim_{u_1,\dots,u_N\to
0\atop{v_1,\dots,v_N\to 0}}
\det_N\bigg[\frac{e^{N(t+u_j+v_k)}}{1-e^{-t-u_j-v_k}}\bigg]\Delta^{-1}(u)
\Delta^{-1}(v).
\end{equation}
It is a Cauchy  determinant; hence,
\begin{equation}\label{hom-lim-res}
f_0(\nu,e^{-t})=\lim_{u_1,\dots,u_N\to
0\atop{v_1,\dots,v_N\to
0}}\frac{\Delta(e^{-u})\Delta(e^{-v})}{\Delta(u)
\Delta(v)}\prod_{j=1}^N
e^{u_j+v_j}\prod\limits_{j,k=1}^N\left(1-e^{-t-u_j-v_k}\right)^{-1}.
\end{equation}
The calculation of the homogeneous limit becomes now trivial and we
obtain
\begin{equation}\label{result_int}
f_0(\nu,w)=(1-w)^{-N^2}.
\end{equation}
\subsection{Non-integer $\nu$}

Here we only give a sketch of the proof. Let us have
$|w|=1$ but $w\ne 1$. Let also $\nu$ satisfy $-1/2 \leq \Re(\nu) \leq 1/2$. The general value of $\nu$
can be reached by analytic continuation.

It is readily seen that
the infinite matrix $V$ in \eqref{inf-det} can be factorized as a product of two matrices:
\begin{equation}\label{det-inf-fact}
f_0(\nu,w)=\det
\left[I-U(\nu)U(-\nu)\right],
\end{equation}
where
\begin{equation}
U_{jk}(\nu)=\frac{\sin \pi\nu}{\pi (j+k-1)}\, w^{(j+k-1)/2}\, \frac{\Gamma(j-\nu)\Gamma(k+\nu)}{\Gamma(j)\Gamma(k)}, \qquad j,k=1,2,\dots .
\end{equation}
Let $T(\nu)$ and $\widetilde{T}(\nu)$ be the following Toeplitz matrices:
\begin{equation}
T_{jk}(\nu)=\frac{w^{(k-j)/2}}{j-k+\nu}\qquad\text{and}\qquad \widetilde{T}_{jk}(\nu)=w^{j-k}\,T_{jk}(\nu).
\end{equation}
Using the following identity for the $\Gamma$-functions \cite{PruBM90},
\begin{equation}\label{Prud}
 \sum_{j=0}^\infty\frac{\Gamma(j-\nu+1)}{j!(j+p)}=
 \frac{\Gamma(p)\Gamma(\nu)\Gamma(1-\nu)}{\Gamma(p+\nu)},
\end{equation}
it is easy to show that
\begin{equation}
U(-\nu)T(-\nu)=\widetilde{T}(\nu)U(\nu)=H(\nu),
\end{equation}
where $H(\nu)$ is a Hankel matrix:
\begin{equation}
H_{jk}(\nu)=\frac{w^{(k+j-1)/2}}{j+k-1+\nu}.
\end{equation}
This means that the determinant representation \eqref{det-inf-fact} can be rewritten in terms of Toeplitz and Hankel matrices:
\begin{equation}\label{det_th}
f_0(\nu,w)=\det
\left[I-\widetilde{T}^{-1}(\nu)H(\nu)H(\nu)T^{-1}(-\nu)\right].
\end{equation}

Now we have to recall some general results for Toeplitz and Hankel matrices. Let $a(z)$ be a smooth function on a unit circle. Its Fourier coefficients are
\begin{equation*}
[a]_n = \frac 1{2\pi}\int\limits_{0}^{2\pi}d\theta\, e^{-in\theta}\,a\big( e^{i\theta}\big).
\end{equation*}
We also define the function $\tilde{a}$ such that $[\tilde{a}]_n=[a]_{-n} $.
 The   Hankel   and Toeplitz operators associated with such a smooth function can be defined as
 \begin{equation*}
 H_{jk}[a]=[a]_{j+k-1},\qquad  T_{jk}[a]=[a]_{j-k}, \qquad\text{with}\quad j,k=1,2,\dots .
\end{equation*}
We will use also the following identity for the Hankel and Toeplitz matrices: for any two functions smooth on the unit circle, it is easy to demonstrate, using basic properties of the Fourier coefficients, that
\begin{equation}\label{toeplitz_prod}
T[ab]=T[a]T[b]+H[a]H[\tilde{b}].
\end{equation}

Assume that the function $a$ admits the Wiener--Hopf factorization $a(z)=a_+(z)a_-(z)$ with unit constant term, where
\begin{equation*}
 a_+ (z)=\exp\left\{\sum_{k=1}^\infty z^k[\ln a]_k\right\},\quad a_- (z)=\exp\left\{\sum_{k=1}^\infty z^{-k}[\ln a]_{-k}\right\}.
\end{equation*}
Then the corresponding Toeplitz matrix can  also be factorized as $T[a]=T[a_-]T[a_+]$.
Moreover, it can be easily shown that, for any function $b$, the following identities hold: $T[ab]=T[a_-]T[a_+b]=T[a_-b]T[a_+]$.

The last result we need was obtained by H. Widom \cite{Wid75} for the determinant of products of Toeplitz matrices. If $a$ and $b$ are smooth and  admit the Wiener--Hopf factorization, the following determinant can be computed:
\begin{equation}\label{Widom}
\det\left[ T^{-1}[a_+]T[a_+b_-]T^{-1}[b_-]\right]=\exp\left\{\sum_{k=1}^\infty k\,[\ln a]_k[\ln b]_{-k}\right\}.
\end{equation}

Now we can apply these results to compute the determinant \eqref{det_th}.
We define the following functions:
\begin{equation}
a(z)= \sum_{n=-\infty}^\infty \,\frac{w^{ n/2}z^n}{n+\nu},\qquad b(z)= \sum_{n=-\infty}^\infty \,\frac{w^{- n/2}z^n}{n-\nu}.
\end{equation}
The corresponding Toeplitz and Hankel operators are exactly the ones appearing in  \eqref{det_th}:
\begin{equation}
H[a]=-H[\tilde{b}]=H(\nu), \qquad T[b]=T(-\nu), \qquad T[a]=\widetilde{T}(\nu).
\end{equation}
It is easy to write an explicit form of these functions. Setting $w^{1/2}= e^{i\psi}$, we obtain
\begin{alignat}{3}
a(e^{i\theta})=&\frac{2i\pi  e^{-i\nu(\theta+\psi)}}{1-e^{2i\pi\nu}},\quad &\text{with}\quad&\theta\in[-\psi,2\pi-\psi],\\
b(e^{i\theta})=&\frac{2i\pi  e^{i\nu(\theta+2\pi-\psi)}}{1-e^{2i\pi\nu}},\quad &\text{with}\quad&\theta\in[-2\pi+\psi,\psi].
\end{alignat}
It is important to note that, although these functions are not smooth on the unit circle, all the results for the Toeplitz and Hankel matrices mentioned  above can nevertheless be used also for such functions \cite{Ehr97}.

Both functions admit the Wiener--Hopf factorization; it is easy to see that
\begin{align}
[\ln a]_n&=\ln\left(\frac\pi{\sin\pi\nu}\right)\delta_{n,0}+\frac\nu n w^{n/2}\left(1-\delta_{n,0}\right), \nonumber\\
[\ln b]_n&=\ln\left(-\frac\pi{\sin\pi\nu}\right)\delta_{n,0}-\frac\nu n w^{n/2}\left(1-\delta_{n,0}\right).\label{wh-ab}
\end{align}

Now we can easily compute the determinant  \eqref{det_th}:
\begin{equation}\label{det_th1}
f_0(\nu,w)=\det
\left[I+{T}^{-1}[a]H[a]H[\tilde{b}]T^{-1}[b]\right]=\det
\left[{T}^{-1}[a]T[ab]]T^{-1}[b]\vphantom{I+{T}^{-1}[a]H[a]H[\tilde{b}]T^{-1}[b]}\right],
\end{equation}
where we used \eqref{toeplitz_prod}. Applying the Wiener--Hopf factorization we rewrite the determinant as
\begin{equation*}
f_0(\nu,w)
=\det\left[ T^{-1}[a_+]T[a_+b_-]T^{-1}[b_-]\vphantom{I+{T}^{-1}[a]H[a]H[\tilde{b}]T^{-1}[b]}\right].
\end{equation*}
Finally, using \eqref{Widom} and \eqref{wh-ab}, we obtain
\begin{equation}\label{result_nonint}
f_0(\nu,w)=\exp\left\{-\sum_{n=1}^\infty \frac{\nu^2}n w^n\right\}=(1-w)^{-\nu^2}.
\end{equation}



\subsection{Summation formula for $\ell\neq 0$}

The general $\ell$ case can be reduced to the case $\ell=0$  by a simple combinatorial trick which we call `background shift'. It is convenient to consider the sum in \eqref{magic formula}
as a sum over all possible excitations over a Dirac sea (all non-positive integers) which can be constructed as either particles (positive integers $p_j$) or holes (non-positive integers $1-h_j$) with a fixed difference between numbers of particles and holes $n_p-n_h=\ell$.  Equivalently, every term of the sum can be considered as an excitation over a shifted Dirac sea (all integers less than or equal to $\ell$) with particles in the positions $\ell+\tilde{p}_k$, $\tilde{p}_k=1,2,\dots$ and holes at $1+\ell-\tilde{h}_k$, $\tilde{h}_k=1,2,\dots$. This second parametrization is convenient because there the number of particles is equal to the number of holes.

Before giving the general case we illustrate the background shift in the simplest case $\ell=1$.
Let us consider the following sum:
\begin{multline}\label{Ext-For-1}
f_1(\nu,w)=\sum_{n=0}^\infty\quad
\sum_{ \substack{ p_1<\cdots< p_{n+1} \\ p_a \in \mathbb{N}^*}   }\quad
\sum_{ \substack{ h_1< \cdots< h_{n} \\ h_a \in \mathbb{N}^*}   }
\prod_{j=1}^{n+1}w^{p_j-1}\prod_{j=1}^n w^{h_j}\\
\times\left(\frac{\sin\pi
\nu}\pi\right)^{2n}\frac{\prod\limits_{j>k}^{n+1}(p_j-p_k)^2\prod\limits_{j>k}^{n}(h_j-h_k)^2}
{\prod\limits_{j=1}^{n+1}\prod\limits_{k=1}^{n}(p_j+h_k-1)^2}
\prod_{k=1}^{n+1}\frac{\Gamma^2(p_k+\nu)}{\Gamma^2(p_k)}
 \prod_{k=1}^{n}\frac{\Gamma^2(h_k-\nu)}{\Gamma^2(h_k)}.
\end{multline}
We take an arbitrary term of this sum with a given $n$ and given positions of particles and holes. We can distinguish two possible situations: $p_1=1$ and $p_1>1$.
In the first case  there is no hole at  point $1$ and we define the positions of particles and holes with respect to the shifted Dirac sea as follows:
\[\tilde{p}_j=p_{j+1}-1,\quad j=1,\dots,n,\qquad\tilde{h}_j=h_{j}+1,\quad j=1,\dots,n.\]
Thus we obtain a configuration with $n$ particles and $n$ holes. Substituting this into \eqref{Ext-For-1}, after
elementary algebra we arrive at
\begin{multline*}
\prod_{j=1}^{n+1}w^{p_j-1}\prod_{j=1}^n w^{h_j}
\left(\frac{\sin\pi
\nu}\pi\right)^{2n}\frac{\prod\limits_{j>k}^{n+1}(p_j-p_k)^2\prod\limits_{j>k}^{n}(h_j-h_k)^2}
{\prod\limits_{j=1}^{n+1}\prod\limits_{k=1}^{n}(p_j+h_k-1)^2}
\prod_{k=1}^{n+1}\frac{\Gamma^2(p_k+\nu)}{\Gamma^2(p_k)}
 \prod_{k=1}^{n}\frac{\Gamma^2(h_k-\nu)}{\Gamma^2(h_k)}\\
 =\Gamma^2(1+\nu)
\prod_{j=1}^{n}w^{\tilde{p}_j-1}\prod_{j=1}^n w^{\tilde{h}_j}
\left(\frac{\sin\pi
\nu}\pi\right)^{2n}\\
\times\left(\det_{n}\,\,\frac1{\tilde{p}_j+\tilde{h}_k-1}\right)^2
\prod_{k=1}^{n}\frac{\Gamma^2(\tilde p_k+\nu+1)\Gamma^2(\tilde h_k-\nu-1)}
{\Gamma^2(\tilde p_k)\Gamma^2(\tilde h_k)}.
\end{multline*}

In the second case ($p_1>1$) there is a hole at point $1$. We set
\[\tilde{p}_j=p_{j}-1,\quad j=1,\dots,n+1,\qquad\tilde{h}_1=1,\quad\tilde{h}_j=h_{j-1}+1,\quad j=2,\dots,n+1,\]
and we obtain a configuration with $n+1$ particles and $n+1$ holes.
Substituting this into \eqref{Ext-For-1}, we obtain
\begin{multline*}
\prod_{j=1}^{n+1}w^{p_j-1}\prod_{j=1}^n w^{h_j}
\left(\frac{\sin\pi
\nu}\pi\right)^{2n}\frac{\prod\limits_{j>k}^{n+1}(p_j-p_k)^2\prod\limits_{j>k}^{n}(h_j-h_k)^2}
{\prod\limits_{j=1}^{n+1}\prod\limits_{k=1}^{n}(p_j+h_k-1)^2}
\prod_{k=1}^{n+1}\frac{\Gamma^2(p_k+\nu)}{\Gamma^2(p_k)}
 \prod_{k=1}^{n}\frac{\Gamma^2(h_k-\nu)}{\Gamma^2(h_k)}\\
 =\Gamma^2(1+\nu)
\prod_{j=1}^{n+1}w^{\tilde{p}_j-1}\prod_{j=1}^{n+1} w^{\tilde{h}_j}
\left(\frac{\sin\pi
\nu}\pi\right)^{2(n+1)}\\
\times\left(\det_{n+1}\,\,\frac1{\tilde{p}_j+\tilde{h}_k-1}\right)^2\quad
\prod_{k=1}^{n+1}\frac{\Gamma^2(\tilde p_k+\nu+1)\Gamma^2(\tilde h_k-\nu-1)}
{\Gamma^2(\tilde p_k)\Gamma^2(\tilde h_k)}.
\end{multline*}
It is easy to see that, up to a general coefficient $\Gamma^2(1+\nu)$, we retrieve exactly the same terms as in the $\ell=0$ case \eqref{Ext-For}. It is also easy to observe that the background shift produces  a one to one correspondence between the configurations of particles and holes in \eqref{Ext-For-1} and in \eqref{Ext-For}, up to the replacement of $\nu$ by $\nu+1$. This leads to a very simple result:
\begin{equation}
f_1(\nu,w)
=\Gamma^{2}(1+\nu)f_0(\nu+1,w).
\label{1_result}
\end{equation}

We can now use this approach in the case of general $\ell$. Let for definiteness $\ell>0$. We define  $f_\ell(\nu,w)$ for $|w|<1$ and arbitrary complex $\nu$
 by the following series:
\begin{multline}\label{Ext-For-l}
f_\ell(\nu,w)=\sum_{n=0}^\infty\quad
\sum_{ \substack{ p_1<\cdots< p_{n+\ell} \\ p_a \in \mathbb{N}^*}   }\quad
\sum_{ \substack{ h_1< \cdots< h_{n} \\ h_a \in \mathbb{N}^*}   }
\prod_{j=1}^{n+\ell}w^{p_j-1}\prod_{j=1}^n w^{h_j}\\
\times\left(\frac{\sin\pi
\nu}\pi\right)^{2n}\frac{\prod\limits_{j>k}^{n+\ell}(p_j-p_k)^2\prod\limits_{j>k}^{n}(h_j-h_k)^2}
{\prod\limits_{j=1}^{n+\ell}\prod\limits_{k=1}^{n}(p_j+h_k-1)^2}
\prod_{k=1}^{n+\ell}\frac{\Gamma^2(p_k+\nu)}{\Gamma^2(p_k)}
 \prod_{k=1}^{n}\frac{\Gamma^2(h_k-\nu)}{\Gamma^2(h_k)}.
\end{multline}
We consider an arbitrary term of this sum with $n+\ell$ parameters $p_k$ and $n$ parameters $h_j$.  We introduce a new equivalent set of parameters $\{\tilde{p},\tilde{h}\}$ as follows.
\begin{itemize}
\item There is a value $m$, $0\le m\le \ell$ such that for $k\le m$ all the parameters $p_k\le\ell$, while for $k>m$ all the parameters $p_k>\ell$. Then for  $k=m+1,\dots,n+\ell$ we define $\tilde{p}_{k-m}=p_k-\ell$.
\item There are $\ell-m$ integers $\ell\ge j_1>j_2>\dots>j_{\ell-m}\ge1$  such that  $p_k\neq j_i$,  $\forall k,i$. Then we define $\tilde{h}_i=\ell-j_i+1$, for $i=1,\dots, \ell-m$.
\item We define $\tilde{h}_{k+\ell-m}=h_k+\ell$, for $k=1,\dots, n$.
\end{itemize}
As in the case $\ell=1$ an advantage of this new set of parameters is the fact that there are the same numbers ($n+\ell-m$) of parameters $\{\tilde{p}\}$ and $\{\tilde{h}\}$. After some computations (similar to those for the $\ell=1$ case but more cumbersome), we obtain
\begin{multline*}
\prod_{j=1}^{n+\ell}w^{p_j-1}\prod_{j=1}^n w^{h_j}
\left(\frac{\sin\pi
\nu}\pi\right)^{2n}\frac{\prod\limits_{j>k}^{n+\ell}(p_j-p_k)^2\prod\limits_{j>k}^{n}(h_j-h_k)^2}
{\prod\limits_{j=1}^{n+\ell}\prod\limits_{k=1}^{n}(p_j+h_k-1)^2}
\prod_{k=1}^{n+\ell}\frac{\Gamma^2(p_k+\nu)}{\Gamma^2(p_k)}
 \prod_{k=1}^{n}\frac{\Gamma^2(h_k-\nu)}{\Gamma^2(h_k)}\\
=w^{\ell(\ell-1)/2}\prod_{j=1}^\ell\Gamma^2(\nu+j)\prod_{j=1}^{n+\ell-m}w^{\tilde{p}_j+\tilde{h}_j-1}\left(\frac{\sin\pi
\nu}\pi\right)^{2(n+\ell-m)}\\
\times\left(\det_{n+\ell-m}\,\,\frac1{\tilde{p}_j+\tilde{h}_k-1}\right)^2\quad
\prod_{k=1}^{n+\ell-m}\frac{\Gamma^2(\tilde p_k+\nu+\ell)\Gamma^2(\tilde h_k-\nu-\ell)}
{\Gamma^2(\tilde p_k)\Gamma^2(\tilde h_k)}.
\end{multline*}
It is easy to note that there is a one to one correspondence between all the possible configurations $\{p,h\}$ and $\{\tilde{p},\tilde{h}\}$. Hence we retrieve the  sum \eqref{Ext-For} in terms of the parameters $\tilde{p}$ and $\tilde{h}$ with a simple prefactor:
\begin{align}
f_\ell(\nu,w)&=w^{\ell(\ell-1)/2}\left(\prod_{j=1}^\ell\Gamma^2(\nu+j)\right)\,f_0(\nu+\ell,w)
\nonumber\\
&=w^{\ell(\ell-1)/2}\left(\prod_{j=1}^\ell\Gamma^2(\nu+j)\right)\,(1-w)^{-(\nu+\ell)^2}.
\label{pos_l_result}
\end{align}

For negative $\ell$ the computation can be performed in a similar way leading to the following result:
\begin{equation}
f_\ell(\nu,w)
=w^{\ell(\ell-1)/2}\left(\prod_{j=1}^{-\ell}\Gamma^{-2}(\nu+1-j)\right)\,(1-w)^{-(\nu+\ell)^2}.
\label{neg_l_result}
\end{equation}
The results for positive and negative values of $\ell$ can be written together using the Barnes $G$-function:
\begin{equation}
f_\ell(\nu,w)
=w^{\ell(\ell-1)/2}\frac{G^2(1+\ell+\nu)}{G^2(1+\nu)}\,(1-w)^{-(\nu+\ell)^2}.
\label{gen_l_result}
\end{equation}
Now setting $w=e^{2\pi ix/L-\varepsilon}$, with $0<x<L$, and taking the limit $\varepsilon\to +0$, we complete the proof of  equation \eqref{magic formula}.



\begin{thebibliography}{99}

%
\bibitem{Bet31} H. Bethe, Zeitschrift f\"ur
Physik {\bf 71} (1931)  205.
%
\bibitem{Hul38} L. Hulth\'en, Arkiv Mat. Astron. Fys. {\bf{26A}} (1938) 1.
%
\bibitem{Orb58} R. Orbach, Phys. Rev.
{\bf 112} (1958)  309.
%
\bibitem{Wal59} L. R. Walker, Phys. Rev.
{\bf 116} (1959)  1089.
%
\bibitem{LieSM61}  E. Lieb, T. Shultz and D. Mattis,
Ann. Phys. {\bf 16} (1961)  407.
%
\bibitem{LieM66}  E. Lieb and D. Mattis (eds.), {\it Mathematical Physics in One Dimension},
New York: Academic Press, 1966.
%
\bibitem{YanY66} C. N. Yang and C. P. Yang, Phys. Rev.
{\bf 150} (1966)  321, 327.
%
\bibitem{FadST79} L. D. Faddeev, E. K. Sklyanin and L. A. Takhtajan,
 Theor. Math. Phys. {\bf 40} (1979) 688.
 %
 \bibitem{Tha81}
H. B. Thacker, Rev. Mod. Phys. {\bf 53} (1981) 253.
 %
 \bibitem{Bax82L}
R.~J. Baxter,
{\it Exactly solved models in statistical mechanics}, London--New York: Academic Press,
   1982.
  %
\bibitem{GauL83} M. Gaudin,
{\it La Fonction d'Onde de Bethe}, Paris: Masson, 1983.
%
%
\bibitem{BogIK93L}V. E. Korepin, N. M. Bogoliubov,
A. G. Izergin, {\it Quantum Inverse Scattering Method and Correlation Functions}, Cambridge: Cambridge Univ.
Press, 1993.
%
\bibitem{FadLH96} L.D. Faddeev, in: Les Houches Lectures {\it Quantum Symmetries}, eds A. Connes
et al, North Holland, (1998) 149.
%
\bibitem{JimM95L}
M.~Jimbo and T.~Miwa,
{\it Algebraic analysis of solvable lattice models}, AMS, 1995.

%
\bibitem{Mcc68}  B. M. McCoy, Phys. Rev.
{\bf 173} (1968)  531.
%
\bibitem{McCTW77}
B. M. McCoy, C. A. Tracy and T. T. Wu,
Phys. Rev. Lett. {\bf 38} (1977) 793.

%
\bibitem{SatMJ78} M. Sato, T. Miwa, M. Jimbo,
Publ. Res. Int. Math. Sci. {\bf 14} (1978) 223; {\bf 15} (1979) 201,
577, 871; {\bf 16} (1980) 531.
%
\bibitem{McCPW81}
B. M. McCoy, J. H. H. Perk and T. T. Wu,
Phys. Rev. Lett. {\bf 46} (1981) 757.
%
\bibitem{KarW78}
M. Karowski and P. Weisz,
Nucl. Phys. B {\bf 139} (1978) 455.

%
\bibitem{Smi92b}
F. A. Smirnov,
{\it Form factors in completely integrable models of quantum field theory},  Adv. Series in Math. Phys.  {\bf 14}, World Scientific, 1992.
%

\bibitem{CarM90}
J. Cardy and G. Mussardo,
Nucl. Phys. B {\bf 340} (1990) 387.

\bibitem{Mus92}
G. Mussardo,
Phys. Rep. {\bf 218} (1992) 215.
%
\bibitem{FriMS90} A. Fring, G. Mussardo and P. Simonetti, Nucl. Phys. B {\bf 393} (1990) 413.
%
\bibitem{KouM93} A. Koubek and G. Mussardo, Phys. Lett. B {\bf 311} (1993) 193.
%
\bibitem{AhnDM93} C. Ahn, G. Delfino and G. Mussardo, Phys. Lett. B {\bf 317} (1993) 573.
%
\bibitem{JimMMN92}
M.~Jimbo, K.~Miki, T.~Miwa and A.~Nakayashiki,
Phys. Lett. A {\bf 168} (1992) 256.
%
\bibitem{JimM96}
M.~Jimbo and T.~Miwa,
J. Phys. A: Math. Gen. {\bf 29}  (1996) 2923.
%

%
\bibitem{Zam91}
A. B. Zamolodchikov,
Nucl. Phys. B {\bf 348} (1991) 619.

%
\bibitem{LukZ97}
S. Lukyanov and A.  Zamolodchikov,
Nucl. Phys. B {\bf 493} (1997) 571.
%
\bibitem{Luk99}
S.~Lukyanov,
Phys. Rev. B {\bf 59} (1999) 11163.
%
\bibitem{LukZ01}
S. Lukyanov and A.  Zamolodchikov,
Nucl. Phys. B {\bf 607} (2001) 437.
%
\bibitem{LukZ01a}
S. Lukyanov,
Nucl. Phys. B {\bf 612} (2001) 391.

%
\bibitem{KitMT99} N. Kitanine, J. M. Maillet and V. Terras,
Nucl. Phys. B  {\bf 554}  (1999) 647.
%
%
\bibitem{KitMT00}
N.~Kitanine, J.~M. Maillet, and V.~Terras,
\newblock Nucl. Phys. B {\bf 567} (2000) 554.
%
%
\bibitem{MaiT00}
J.~M. Maillet and V.~Terras, Nucl. Phys. B {\bf 575} (2000) 627.
%
%
\bibitem{KitMST02a}N. Kitanine, J. M. Maillet, N. A. Slavnov and V. Terras,  Nucl. Phys. B {\bf 641} (2002) 487.
%

%
\bibitem{KitMST05a}N. Kitanine, J. M. Maillet, N. A. Slavnov and V. Terras,
Nucl. Phys. B {\bf 712} (2005) 600.
%
%
\bibitem{KitKMST09b} N. Kitanine, K.~K. Kozlowski, J. M. Maillet, N. A. Slavnov and V. Terras,
J. Stat. Mech. (2009) P04003.



%
\bibitem{GohKS04} F. G\"ohmann, A. Kl\"umper and A. Seel,
J. Phys. A: Math. Gen. {\bf 37} (2004) 7625.
%

%
\bibitem{GohKS05} F. G\"ohmann, A. Kl\"umper and A. Seel,
J. Phys. A: Math. Gen. {\bf 38} (2005) 1833.
%

\bibitem{BooG09}
H. Boos and F. G\"ohmann,
J. Phys. A: Math. Gen. {\bf 37} (2009) 315001.
%
\bibitem{BooJMST07}
H. Boos, M. Jimbo, T.  Miwa, F. Smirnov and Y. Takeyama,
Comm. Math. Phys. {\bf 272} (2007) 263.
%
\bibitem{BooJMS09}
H. Boos, M. Jimbo, T.  Miwa and F. Smirnov,
Comm. Math. Phys. {\bf 299} (2009) 825.

%
\bibitem{JimMS11}
M. Jimbo, T.  Miwa and F. Smirnov,
Lett. Math. Phys. {\bf 96} (2011) 325.

%
\bibitem{JimMS11a}
M. Jimbo, T.  Miwa and F. Smirnov,
Nucl. Phys.  B {\bf 852} (2011) 390.
%
\bibitem{CauM05}
J.~S. Caux and J.~M. Maillet,
Phys. Rev. Lett. {\bf 95} (2005) 077201.
%


%
\bibitem{LutP75} A. Luther and I. Peschel,  Phys. Rev. B
{\bf 12} (1975) 3908.
%
\bibitem{Hal80} F. D. M. Haldane, Phys. Rev. Lett. {\bf 45} (1980)
1358.
%
\bibitem{Hal81a} F. D. M. Haldane, Phys. Lett. A {\bf 81} (1981) 153.
%
\bibitem{Hal81b} F. D. M. Haldane, J. Phys. C: Solid State Phys. {\bf 14} (1981) 2585.
%
\bibitem{BelPZ84}
A. A. Belavin,  A. M. Polyakov and A. B. Zamolodchikov,
Nucl. Phys. B {\bf 241} (1984) 333.
%
%
\bibitem{Aff85} I. Affleck,  Phys. Rev. Lett. {\bf 55} (1985) 1355.
%
\bibitem{BloCN86}  H. W. J. Bl\"ote, J. L. Cardy and M. P. Nightingale,
Phys. Rev. Lett. {\bf 56} (1986) 742.
%
\bibitem{Car84} J. L. Cardy,  J. Phys. A: Math. Gen. {\bf 17} (1984) L385.
%
\bibitem{Car86}  J. L. Cardy,
Nucl. Phys. B {\bf 270} (1986) 186.
%
\bibitem{AlcBB88}
F. C. Alcaraz, M. N. Barber and M. T. Batchelor,
Ann. Phys. {\bf 182} (1988) 280.
%
\bibitem{WoyE87}
F. Woynarovich and H. P. Eckle,
J. Phys. A: Math. Gen. {\bf 20} (1987) L97.
%
%
\bibitem{ColIKT93} F. Colomo, A. G. Izergin, V. E. Korepin, V. Tognetti,
Theor. Math. Phys.  {\bf 94} (1993) 11.
%
\bibitem{KorS90} V. E.~Korepin and N. A.~Slavnov, Comm. Math. Phys. {\bf 129} (1990), 103.
%
\bibitem{Sla90} N. A.~Slavnov, Theor. Math. Phys. {\bf 82} (1990) 273.
%
\bibitem{KitKMST09c} N. Kitanine, K. K. Kozlowski, J. M. Maillet, N. A. Slavnov and V. Terras,
J. Math. Phys. {\bf 50} (2009) 095209.
%
\bibitem{KitKMST11a} N. Kitanine, K.~K. Kozlowski, J. M. Maillet, N. A. Slavnov and V. Terras,
J. Stat. Mech. (2011) P05028.
%
\bibitem{Oot04} T. Oota, J. Phys. A: Math. Gen. {\bf 37} (2004) 441.
%
\bibitem{DegM09} T. Deguchi and  C. Matsui,
Nucl. Phys. B {\bf 814} (2009) 409.
%
\bibitem{KorS99} V.E.~Korepin and N.A.~Slavnov, Int. J. Mod. Phys. B {\bf 13} (1999) 2933.
%
%
\bibitem{CauHM05}
J.~S. Caux, R.~Hagemans and J.~M. Maillet,
J. Stat. Mech. (2005) P09003.
%
\bibitem{PerSCHMWA06}
R.~G. Pereira, J. Sirker J, J.~S. Caux, R. Hagemans, J.~M. Maillet, S.~R. White and I.~Affleck,
Phys. Rev. Lett. {\bf 96} (2006) 257202.
%
\bibitem{PerSCHMWA07}
R.~G. Pereira, J. Sirker J, J.~S. Caux, R. Hagemans, J.~M. Maillet, S.~R. White and I.~Affleck,
J. Stat. Mech. (2007) P08022.
%
\bibitem{CauCS07}
J.~S. Caux, P. Calabrese and N. A. Slavnov,
J. Stat. Mech. (2007) P01008.
%
\bibitem{KitKMST09a}
N. Kitanine, K.~K. Kozlowski, J. M. Maillet, N. A. Slavnov and V. Terras,
Comm. Math. Phys. {\bf 291} (2009) 691.
%
\bibitem{KozMS11a}
K.K. Kozlowski, J.M. Maillet, N.A. Slavnov,
J. Stat. Mech. (2011) P03018.


%
\bibitem{KozT11} K. K.~Kozlowski and V.~Terras, J. Stat. Mech. (2011) P09013.
%
\bibitem{Koz11} K.K.~Kozlowski, \textit{Large-distance and long-time asymptotic behavior
			of the reduced denisty matrix in the non-linear Schr\"{o}dinger model}, math-ph:11011626.

%
\bibitem{LesSS96}
F. Lesage, H. Saleur and S. Skorik,
Nucl. Phys. B {\bf 474} (1996) 602.
\bibitem{LesS97}
F. Lesage and H. Saleur,
J. Phys. A: Math. Gen. {\bf 30} (1997) L457.

\bibitem{LesKS2003}
A. Koutouza, F. Lesage and H. Saleur,
Phys. Rev. B {\bf 68} (2003) 115422.
%
%
\bibitem{KojKS1997} T.~Kojima, V.E.~Korepin and N.A.~Slavnov,
 Comm. Math. Phys.  {\bf 188} (1997) 657.
%
%
%
\bibitem{KerOV93} S. Kerov, G.  Olshanski and A. Vershik,  Comptes Rend. Acad. Sci. Paris, Ser. I  {\bf 316} (1993) 773.
%
\bibitem{Ker2000} S. Kerov, Funct. Anal. Appl. {\bf 34} (2000) 41.

%
\bibitem{BorO2000} A. Borodin, and G. Olshanski,
Electron. J. Combin. {\bf 7} (2000) R28.
%
\bibitem{BorO2000a} A. Borodin and G. Olshanski,
Comm. Math. Phys. {\bf 211} (2000) 335.
%
\bibitem{Oko01}  A. Okounkov,  In: {\it Random matrix models and
their applications},   Math. Sci. Res. Inst. Publ., {\bf 40},
Cambridge Univ. Press,  (2001) 407.
%
\bibitem{BorO2001}
A. Borodin and G. Olshanski,
 In: {\it Random matrix models and their applications},
(P. M. Bleher and A. R. Its, eds). MSRI Publications {\bf 40}, Cambridge Univ. Press, (2001)
 71Ð94.
%
%
\bibitem{Ols2003}
G.  Olshanski,
In: {\it The orbit method in geometry and physics: in honor of A. A. Kirillov},
(C. Duval, L. Guieu, V. Ovsienko, eds), Progress in Math. {\bf 213} Birkhauser, 2003.
%
\bibitem{KozMS11b} K.K. Kozlowski, J.M. Maillet, N.A. Slavnov,
 J. Stat. Mech. 1103:P03019 (2011)
%
\bibitem{ShaGCI11}
A. Shashi, L.I. Glazman, J.S. Caux, A. Imambekov, Phys. Rev. B. {\bf 84} 045408 (2011)
%
%
\bibitem{IzeK85} A. G. Izergin and V. E. Korepin, Comm. Math. Phys. {\bf 99} (1985) 271.
%
\bibitem{TarV95} V. Tarasov and A. Varchenko, Int. Math. Res. Notices {\bf 13} (1995) 637.
%
%
\bibitem{Wid75}
H. Widom,
Proc. AMS {\bf 50} (1975) 167.
%
\bibitem{Ehr97}
T. Ehrhardt,
{\it Toeplitz determinants with several Fisher-Hartwig singularities},
Ph.D. thesis, Fakult\"at f\"ur Mathematik der Technischen Universit\"at Chemnitz,
Chemnitz, Germany, 1997.
%
\bibitem{PruBM90} A.P. Prudnikov, Yu.A. Brychkov and O.I. Marichev,  {\it Integrals and Series, Vol. 2: Special Functions}, New York: Gordon and Breach, 1990.






\end{thebibliography}
\end{document}